\documentclass[twocolumn]{aastex62}

\hypersetup{linkcolor=blue,citecolor=blue,filecolor=blue,urlcolor=blue}

\graphicspath{{./}{figures/}}

\received{XXX}
\revised{XXX}
\accepted{XXX}
\submitjournal{ApJ}

\shorttitle{Setting the Stage for Cosmic Chronometers I}
\shortauthors{Moresco M. et al.}

\begin{document}
\title{Setting the Stage for Cosmic Chronometers. I.\\ Assessing the Impact of Young Stellar Populations on Hubble Parameter Measurements}

\author{Michele Moresco}
\affil{Dipartimento di Fisica e Astronomia, Universit\`a di Bologna, Via Gobetti 93/2, I-40129, Bologna, Italy}
\affiliation{INAF - Osservatorio di Astrofisica e Scienza dello Spazio di Bologna, via Gobetti 93/3, I-40129 Bologna, Italy}

\author{Raul Jimenez}
\affiliation{ICC, Instituto de Ciencias del Cosmos, University de Barcelona, UB, Marti i Franques 1, E-08028, Barcelona, Spain}
\affiliation{ICREA, Pg. Lluis Companys 23, E-08010 Barcelona, Spain}

\author{Licia Verde}
\affiliation{ICC, Instituto de Ciencias del Cosmos, University de Barcelona, UB, Marti i Franques 1, E-08028, Barcelona, Spain}
\affiliation{ICREA, Pg. Lluis Companys 23, E-08010 Barcelona, Spain}

\author{Lucia Pozzetti}
\affiliation{INAF - Osservatorio di Astrofisica e Scienza dello Spazio di Bologna, via Gobetti 93/3, I-40129 Bologna, Italy}

\author{Andrea Cimatti}
\affiliation{Dipartimento di Fisica e Astronomia, Universit\`a di Bologna, Via Gobetti 93/2, I-40129, Bologna, Italy}
\affiliation{INAF - Osservatorio Astrofisico di Arcetri, Largo E. Fermi 5, I-50125 Firenze, Italy}

\author{Annalisa Citro}
\affiliation{Dipartimento di Fisica e Astronomia, Universit\`a di Bologna, Via Gobetti 93/2, I-40129, Bologna, Italy}
\affiliation{INAF - Osservatorio di Astrofisica e Scienza dello Spazio di Bologna, via Gobetti 93/3, I-40129 Bologna, Italy}
\affiliation{Center for Gravitation, Cosmology and Astrophysics, Department of Physics, University of Wisconsin-Milwaukee, 3135 N. Maryland Avenue, Milwaukee, WI 53211, USA}

\correspondingauthor{Michele Moresco}
\email{michele.moresco@unibo.it}



\begin{abstract}
The expansion history of the Universe can be constrained in a cosmology-independent  way by measuring the differential age evolution of cosmic chronometers. This yields a measurement of the Hubble parameter $H(z)$ as a function of redshift. The most reliable cosmic chronometers known so far are extremely massive and passively evolving galaxies. Age-dating these galaxies is, however, a difficult task, and even a small contribution of an underlying young stellar population could, in principle, affect the age estimate and its cosmological interpretation. We present several spectral indicators to detect, quantify and constrain such contamination in old galaxies, and study how their combination can be used to maximize the purity of cosmic chronometers selection. In particular, we analyze the \ion{Ca}{2} H/K ratio, the presence (or absence) of H$\alpha$ and [OII] emission lines, higher order Balmer absorption lines, and UV flux; each indicator is especially sensitive to a particular age range, allowing us to detect young components ranging between 10 Myr and 1 Gyr.
The combination of these indicators minimizes the contamination to a level below 1\% in the case of ideal data. More importantly, it offers a way to control the systematic error on $H(z)$ as a function of the contamination by young stellar 
populations. We show that for our previous measurements of the Hubble parameter, the possible bias induced by the presence of a younger component is well below the current errors. We envision that these indicators will be instrumental in paving the road for a robust and reliable dating of the old population and its cosmological interpretation.
\end{abstract}


\section{Introduction}

The cosmic chronometers (CC) approach, initially proposed by
\citet{jimenez2002}, can provide cosmology-independent measurements of the expansion
rate of the Universe: the Hubble parameter $H(z)$.

It is difficult to overstate the importance of having independent methods to 
measure the same cosmological parameter. First of all, different methods are 
affected by different systematics offering therefore a test for robustness of the 
measurement. At a more general level, different methods that rely on different 
physics offer powerful consistency checks of the underlying model and can even 
be used to constrain or explore new physics. A case in point is the measurement 
of the Hubble constant $H_0$. 
There is a current discrepancy (at the $\sim 3-4 \sigma$ level) between the 
expansion history measured by the local cosmic ladder method at $z \sim 0$ 
\citep{riess2018a} and the one inferred from Cosmic Microwave 
Background (CMB) observations by assuming the Lambda Cold Dark Matter ($\Lambda$CDM) 
model \citep{planck2016}. While the discrepancy level does not yet reach the 
``golden" discovery threshold, it has attracted attention and has been 
scrutinized closely. There are two solutions to this discrepancy: either it is 
due to systematic errors in either (or both) methods, or it could be the signature 
of new physics and the first sign that the standard minimal $\Lambda$CDM 
cosmological model is not adequate to describe the Universe \citep[for a detailed 
analysis of the tension see e.g.,][and refs therein]{verde2013,bernal2016}. 

The basic idea of the CC method is that the Hubble parameter is related to the 
scale factor $a$ and the differential redshift-time relation ($dt/dz\simeq 
\Delta t/\Delta z$) as:
\begin{equation}
H(z) = \frac{\dot a}{a} = \frac{-1}{1+z} \frac{\Delta z}{\Delta t}
\label{eq:Hz}
\end{equation}
assuming a Friedman-Robertson-Walker (FRW) metric. With this minimal 
assumption, if the differential redshift-time relation could be measured, then 
one could measure the Universe's expansion history in a cosmology-independent 
fashion, therefore testing directly the cosmological model. Measuring 
redshifts, and thus $\Delta z$ (e.g., between two galaxies), is 
straightforward with spectroscopic observations. However, 
measuring $\Delta t$ is much more challenging. It requires 
that standard clocks exist throughout the Universe, and that they can be read. 
The requirements on this are less stringent than one could 
naively imagine since what is needed are {\em differential} ages not absolute 
ages: systematic effects that result in a constant offset on the age determination 
gets canceled out in the differential measurement.
This is the origin of the idea of cosmic chronometers, and that stellar 
evolution could provide such standard clocks \citep{jimenez2002}.
If an old and passively evolving stellar population could be found across a 
range of redshifts, such as its stars had all formed 
synchronously, and evolved passively since, that would be suitable to be our CC
\citep[others clocks have been proposed, e.g.,][but will not be discussed here]
{daly2008,diaferio2011,lavaux2012}.
Main sequence stars burn H into He via the $pp-$chain. Empirical 
evidence for this process has been provided by the solar neutrino experiments 
that measure the neutrino flux from our Sun.
Once the chemical composition of an ensemble of stars composing a simple stellar population is known, one can predict its age with high accuracy (in particular if it is possible to resolve them).
Indeed, stellar dating has been and is used 
now to calculate, independently of cosmology, the age of Universe by dating 
globular clusters \citep[e.g.,][]{jimenez1995,marin2009}, being this one of the 
first hints for a cosmological constant. Relative ages of globular 
clusters are computed accurately to the \% level \citep{marin2009}.
More recently \citep{moresco2012}, a different approach has been suggested that does not directly rely on the estimate of the age of a stellar population, but instead uses a direct spectroscopic observable (the 4000~\AA~ break) known to be linearly related (at fixed metallicity) with the age of the stellar population. In this way, it is possible to rewrite Eq. \ref{eq:Hz} in the form:
\begin{equation}
H(z)=\frac{-A}{(1+z)} \frac{dz}{dD4000}\;,
\label{eq:Hz_Dn}
\end{equation}
with the advantage of decoupling systematic and statistical effects
and of being more robust against the models assumed to calibrate the relation.

Unfortunately, we not do see individual stars or resolved stellar populations 
like globular clusters beyond our Local Group; to find cosmic 
chronometers we have to rely on dating unresolved stellar populations 
\citep{tinsley1968}. 
With the advent of high-resolution spectroscopy in $10$ $m$-class telescopes and 
more accurate stellar models and fitting methods 
\citep[see e.g.][]{reichardt2001,tojeiro2007,choi2014,citro2016,chevallard2016}, 
it is now possible to estimate the ages of the stellar populations of galaxies 
with increasing accuracy. This has been done for large samples like Sloan Digital 
Sky Survey (SDSS), VIMOS and others or targeted small samples of galaxies. If the 
old stellar population of these galaxies (or an identifiable sample 
of them across different redshifts) were synchronized, these could be CC.
\footnote{In principle, any galaxy could be used 
as a CC if the old stellar population is dominant and can be disentangled from the 
young one.} 
There is considerable empirical evidence \citep{dunlop1996,spinrad1997,cowie1999,heavens2004,thomas2005,treu2005,panter2007,citro2016}
for the existence of a population of galaxies, that has formed its stellar 
population very early ($\sim 1$ Gyr after the Big Bang), has finished its 
major star formation episode by $z\gtrsim 2$, and that since then has been 
evolving passively with only very minor ``frosting'', if any at all. These galaxies 
tend to be very massive and harbored in the highest-density regions of galaxy 
clusters. 
In theory, a young stellar component, even if not dominant 
in terms of stellar mass of the global population, can influence the spectrum, and 
potentially bias the measurements. Hence these objects have to be selected 
accurately to be CC and to estimate their old stars age correctly. 
In the last decade, the CC approach has been applied to several dataset of 
increasing size to cover a wider range in redshift thus providing a more 
detailed picture of how $H$ changes as a function of redshift 
\citep{jimenez2003,simon2005,stern2010,moresco2012,zhang2014,moresco2015,
moresco2016}. These measurements yield the expansion history directly 
and independently of the adopted cosmological model; they do not rely on a 
CMB-calibrated ``ruler" as, e.g., Baryon Acoustic Oscillations (BAO) do. This 
makes the CC method so attractive for model testing.

The main aim of this paper is twofold: {\it i)} quantify the impact on the Hubble parameter, estimated with the cosmic chronometer approach, of a possible contamination from a young component and provide its corresponding covariance matrix, as a component of the total error budget; {it ii)} provide a clear set of indicators and a selection work-flow that can be used to select a pure sample of passively evolving galaxies, minimizing (with respect to other approaches adopting only one or few criteria) the possible contamination from a young component.
This paper is organized as follows. In Sec. \ref{sec:problem} we pedagogically 
explain the challenges of the CC approach, and review approaches in the 
literature, in Sec. \ref{sec:method} we describe the method used to explore the 
presence of a young component underlying the spectrum of an older, passively 
evolving population; in Sect. \ref{sec:youngcomp} we explore where the signal of a 
young contribution is localized in the observed spectrum, identifying different 
tracers to discriminate such component at different ages, namely the 
\ion{Ca}{2}$H/K$ ratio, the presence of [OII] or H$\alpha$ emission lines, the UV 
flux, and the presence of strong absorption Balmer lines; in Sect. 
\ref{sec:D4000}, we dissect how such contributions could potentially affect the 
measurement of $D4000$, and hence of Hubble parameter $H(z)$; in Sect. 
\ref{sec:optsel} we present the recipe for an optimal selection of 
cosmic chronometers, based on the results of the previous sections; finally, 
in Sect. \ref{sec:concl}, we draw our conclusions.

This paper is the first of a series of papers in which we will revisit the main 
systematic effects related to the cosmic chronometers method. In paper II we will 
address the dependence of the results on the assumed stellar population synthesis models 
using state-of-the-art models, and in paper III we will discuss in further details the 
issue of metallicity, and how to improve in its measurement.

\section{The problem}
\label{sec:problem}
The CC method relies on accurate {\em differential} dating of 
integrated stellar populations, which is very challenging, and one needs to be 
very aware of the systematics that can impact the parameter extraction from the 
integrated stellar light. First, the integrated light is the convolution of many 
individual bursts of star formation, each with its own age and metallicity. These 
get combined to produce the integrated light of a galaxy. While the {\it direct} 
problem, to predict the integrated light given a star formation law, is 
straightforward, the {\it inverse} problem is much harder. 

In mathematical terms: single stellar populations (SSPs) are the building blocks 
of any arbitrarily complex population since the latter can be computed as a 
sum of SSPs, once the star formation rate is provided. In other words, the 
luminosity of a stellar population of age $t_0$ (since the beginning of star 
formation) can be written as:
\begin{equation}
L_{\lambda} (t_0) = \int_0^{t_{0}} \int_{Z_i}^{Z_f} L_{\lambda}^{SSP} (Z, t_0-t) dZ dt,
\label{eq:ssp}
\end{equation}
where the luminosity of the SSP is:
\begin{equation}
L_{\lambda}^{SSP} (Z, t_0-t) = \int_{M_u}^{M_d} SF(Z,M,t) l_{\lambda} (Z, M, t_0-t) dM
\end{equation}
and $l_{\lambda} (Z,M,t_0-t)$ is the luminosity of a star of mass $M$, 
metallicity $Z$ and age $t_0-t$, $Z_i$ and $Z_f$ are the initial and final 
metallicities, $M_d$ and $M_u$ are the smallest and largest stellar masses in 
the population and $SF(Z,M,t)$ is the star formation rate at the time $t$.
It is worth emphasising that $l_{\lambda} (Z, M, t_0-t)$ can be computed 
accurately from first principles of stellar evolution.
This is the {\it direct} problem, which is well under control. 

But it is easy to see from Eq.~\ref{eq:ssp} that the inversion process needed to 
recover the age of underlying SSPs of a galaxy may fail. It might happen that 
the integrated light will not have enough information to permit a full 
inversion of Eq.~\ref{eq:ssp}. Fortunately, this inversion problem was addressed 
formally in ~\citet{tojeiro2007} \citep[but see also][]
{connolly1995,yip2004,wild2007,chen2012,yip2014} by effectively letting the data decide 
how much information could actually be recovered. Given an amount of data and S/N 
ratio over a given spectral range, the above procedure returns a coarser or finer 
reconstructed star-formation history, with corresponding uncertainty, depending on 
the information content of the data. Of course, other approaches can be 
exploited to estimate the uncertainty on the star-formation history reconstruction,
such as regularization \citep{cappellari2004} and bootstrap \citep{walcher2015}. The 
specific choice of technique is not central to our argument here.

The main systematics that could potentially affects the CC method have 
been extensively studied. These are: \emph{(i)} the dependence on 
stellar metallicity estimate, \emph{(ii)} the reliance on stellar population 
synthesis (SPS) models, \emph{(iii)} the progenitor bias, and \emph{(iv)} the 
presence of an underlying young component. For points \emph{(i)}, \emph{(ii)} and 
\emph{(iii)}, we refer to \citet{moresco2012,moresco2016}, where a detailed 
analysis on these issues have been presented.
The robustness and consistency of CC relative ages estimation has been studied 
before, e.g., by \citet{crawford2010} and \citet{liu2016}. These authors find that 
$H(z)$ can be recovered with 3\% error at intermediate redshifts keeping 
systematic errors safely below the statistical ones. Here we concentrate on point 
\emph{(iv)}.

It is worth recalling the different approaches taken in the literature, 
which have evolved with the size and quality of the data sets.
In \citet{jimenez2003} the age-determination was done by reconstructing the 
so-called red envelope \citep{butcher1983,lilly1985,connell1988}.
Full-spectral fitting assuming a single age was used to age-date a sample of 
simulated galaxies where a young population was superimposed to an old one. 
It is shown \citep[see Fig.5 of][]{jimenez2003} that even without an accurate 
pre-selection of target galaxies, a random burst of recent star formation (a 
young population) in a mostly dominant old population does not affect the 
recovered age-redshift relation obtained from the red envelope, provided there 
are enough galaxies to populate the red envelope. This technique was 
successfully applied to a densely sampled galaxy survey (SDSS).
Alternatively, it is possible to proceed with a more accurate selection of 
passively evolving galaxies, studying their mean evolution as a function of 
cosmic time. Both approaches are viable, each one having its own advantages 
and drawbacks. In \citet{moresco2012}, both have been explored, and it was 
found that the limit of the red envelope approach is that it is highly dependent 
on statistics, needing a large dataset to properly sample the extreme of the 
distribution (e.g. the 5\% or the 1\% oldest galaxies); on the contrary, the 
other approach needs a better control in the sample selection, but it is more 
stable, providing at the end smaller errors \citep[as shown in][]{moresco2012}.

In \citet{stern2010} full spectral fitting with an extended star formation 
history was used and the wavelength coverage required to recover the age-redshift 
relation from simulations was explored (see Fig. 1, 2 and 3) as well 
In \citet{moresco2016} a different approach was taken. A new estimator ($D4000$) 
for the galaxy age was adopted which is robust provided it is applied to a 
pre-selected sample of passively evolving galaxies (CC). It is also shown how 
to recover the metallicity and star formation histories of galaxies and how this 
information can be used to robustly select such a sample of galaxies suitable to 
be CC.

We live in an era of massive spectroscopic surveys which will
bring a treasure trove of galaxies and could further increase the precision 
and accuracy of the CC method. This offers the prospect to provide more precise 
cosmology-independent measurements of $H(z)$ with the potential of new physics 
discoveries. Passively evolving galaxies will not 
be specifically targeted, and spectra will cover the range in the rest-frame 
near-UV to near-IR. Here, we study how to select CC 
from features in the optical rest-frame spectrum and how to eliminate the 
contamination of an underlying star forming population.

In particular, we present several spectral indicators that facilitate to 
disentangle and discriminate the presence of a star-forming population 
contribution contaminating passively evolving galaxies, further strengthening 
the selection criteria of optimal cosmic chronometers, and thus paving the road 
for a robust and reliable dating of the old population. We then present a 
complete ``recipe" to select and age-date cosmic chronometers. 

\begin{figure*}[t!]
  \centering
  \includegraphics[width=\linewidth]{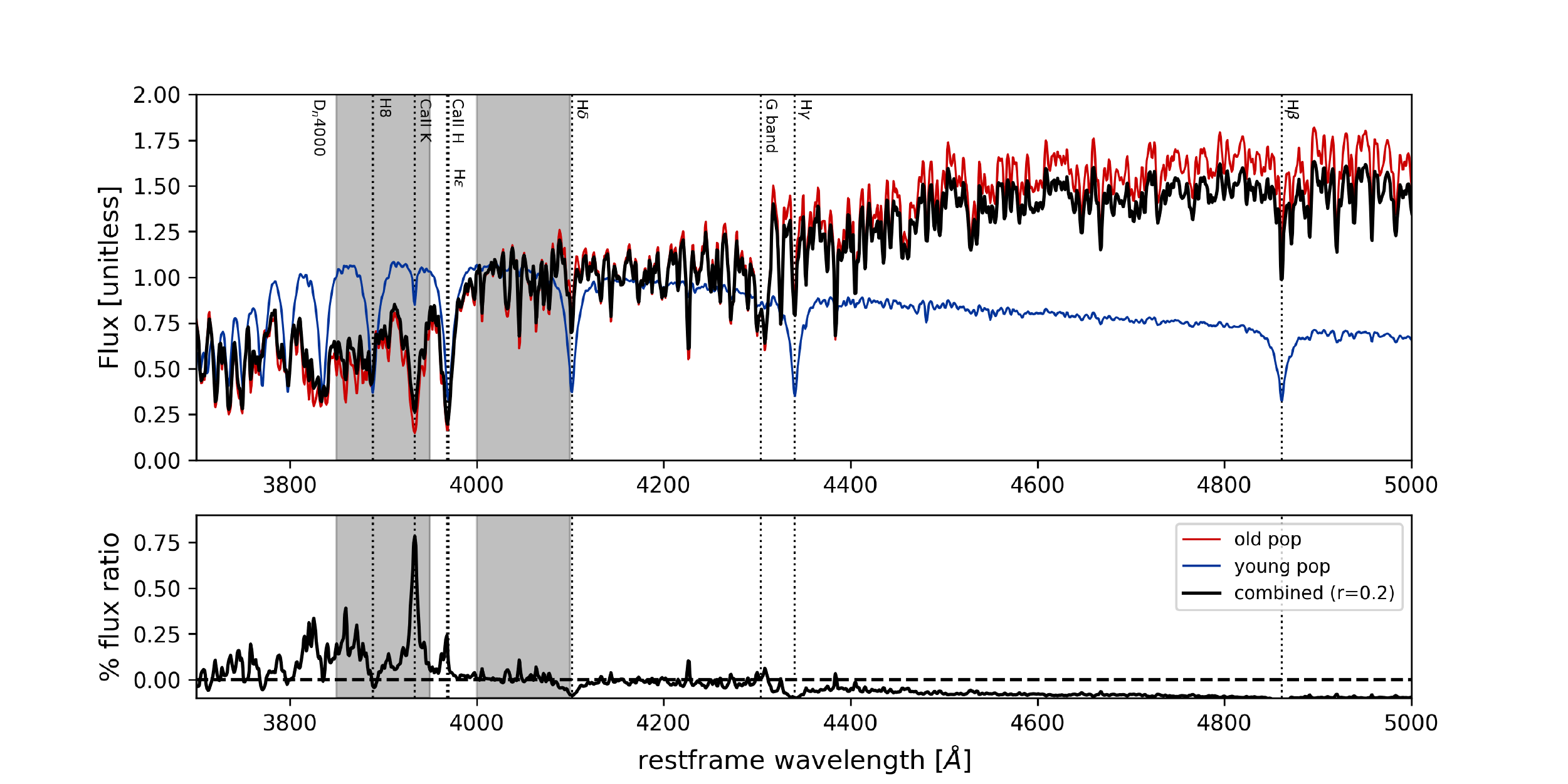} \hfill
  \caption{Synthetic spectra. In the upper panel it is shown, for illustrative purposes, the synthetic spectra of an old component (red line, 8 Gyr), of a young component (blue line, 0.1 Gyr), and their combination, with a fraction $r=0.2$ (black line). All the spectra have been generated with BC16 models assuming a MILES library, Chabrier IMF and solar metallicity. The lower panel show the percentage difference between the old and combined spectrum (namely, $r(\lambda)-1$); as it can be seen, the larger impact is in correspondence to \ion{Ca}{2} K line, due to the onset of the \ion{Ca}{2} $H/K$ inversion.}
     \label{fig:spectrum}
\end{figure*}

\section{Method}
\label{sec:method}

To study the impact of a young component on the spectrum of an older, 
passively evolving galaxy, we build synthetic spectra composed of a mixture of a 
younger and an older population. We explore a variety of different 
models, to assess the dependence of our results on the SPS model assumed. 
We create the library of synthetic spectra using the 
\citet{maraston2011} models (hereafter M11), the updated 2016 \citet{bruzual2003} 
models (hereafter BC16), and the \citet{vazdekis2016} models. All models are based 
on the latest MILES stellar library \citep{falcon2011}, but encompass different 
recipes and ingredients to construct the synthetic spectra \citep[for further 
discussion, see e.g.][]{maraston2006,maraston2011}. We also decide to choose the 
Chabrier initial mass function \citep[IMF,][]{chabrier2003}, but as demonstrated 
in \citet{moresco2012} the impact on the $D4000$ (the main feature analyzed in 
this work) of the adopted IMF is negligible. 

Currently, cosmic chronometer data have been exploited in the redshift range $0.15<z<2$, and we therefore want to span a grid of ages sensible for this entire redshift range: we thus consider for the old component a solar metallicity, exploring the ages=[2, 4, 6, 8, 10, 12] Gyr, and for the young component several values ranging from 0.1 Gyr to 1 Gyr with stellar metallicity in the range $0.4<Z/Z_{\odot}<2.5$. The choice of the ages for the young component is motivated by the fact that in previous CC analysis the effect of an extended SFH has been already taken into account (e.g. a delayed exponential SFH, compatible with the colors and spectra of the data), while here we want to explore the possible bias due to a much more recent burst of star formation.

The fraction of the young population flux contribution with respect to the old population 
is parameterized by $r(\lambda)=F_{young}/F_{old}$.
This definition of $r$ depends on the bandpass used to determine $F_{young}$ with
respect to $F_{old}$. Here we adopt two different wavelength ranges 
(representative of the two $D_{n}4000$ band-passes), namely [4000-4100] ~\AA~ (for 
$r$) and [3850-3950]~\AA~ (for $r'$), to explore the impact of the range of 
normalization on the results. 

In Fig. \ref{fig:spectrum} we show, for illustrative purposes, the
spectra obtained at 8 Gyr for the old population, 0.1 Gyr for the young 
component, and their combination with a fraction $r=0.2$ normalized at $[4000-
4100]$~\AA. From the plots, all the absorption lines
typical of the two components are evident. In particular, the
older spectrum is characterized by a strong $D4000$, significant
\ion{Ca}{2} H and K lines and the G-band feature at 4304~\AA, while
the younger spectrum is dominated by strong Balmer lines, namely
H$\beta$, H$\gamma$, H$\delta$, H$\epsilon$.
In the lower plot is presented an example of the percentage 
difference between the spectrum of the old population and the combined spectrum 
(i.e. $r(\lambda)-1$), showing how this ratio changes as a function of the 
wavelength.

\begin{figure*}[t!]
 \centering
  \includegraphics[width=\columnwidth]{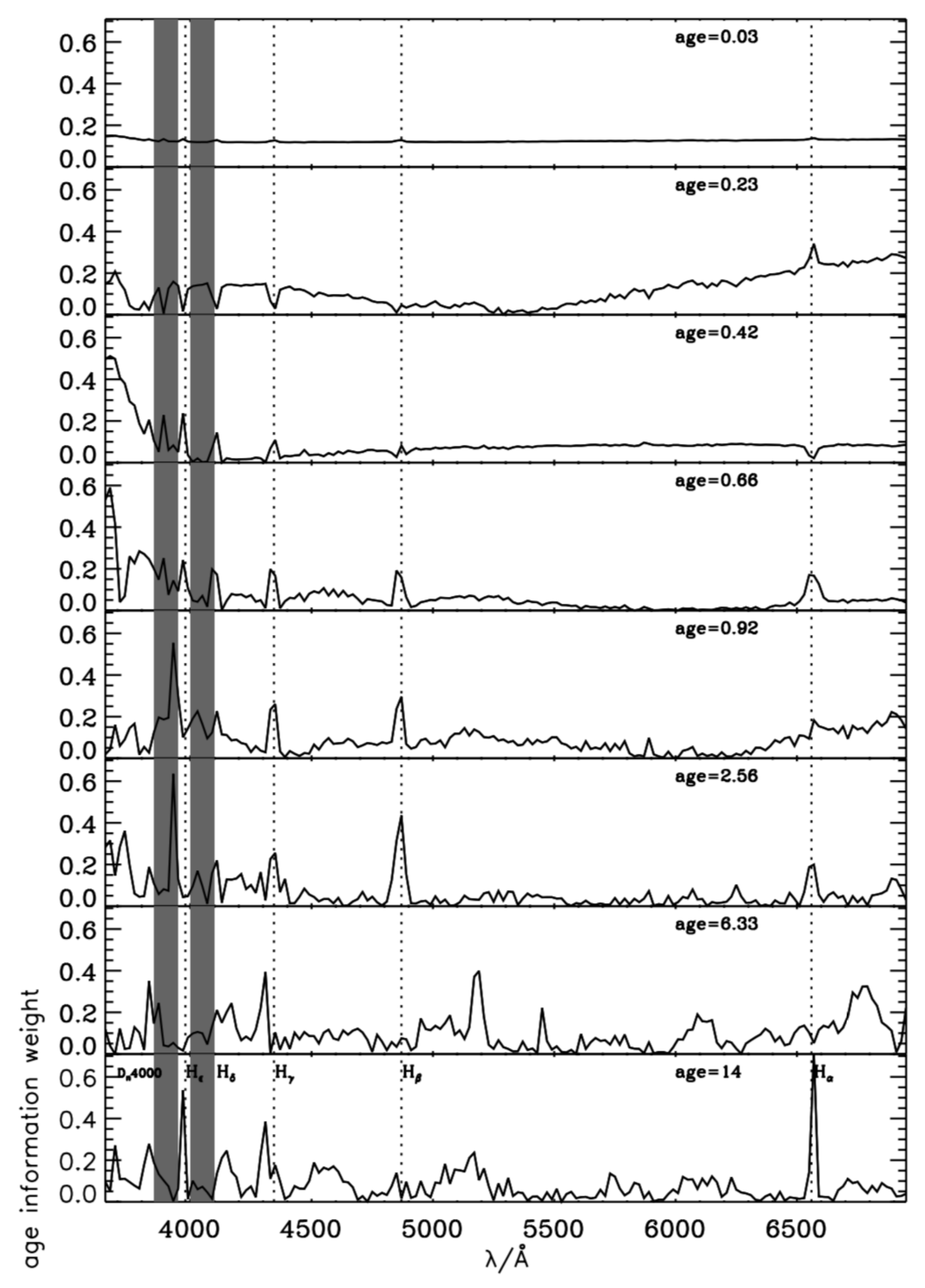}
  \includegraphics[width=\columnwidth]{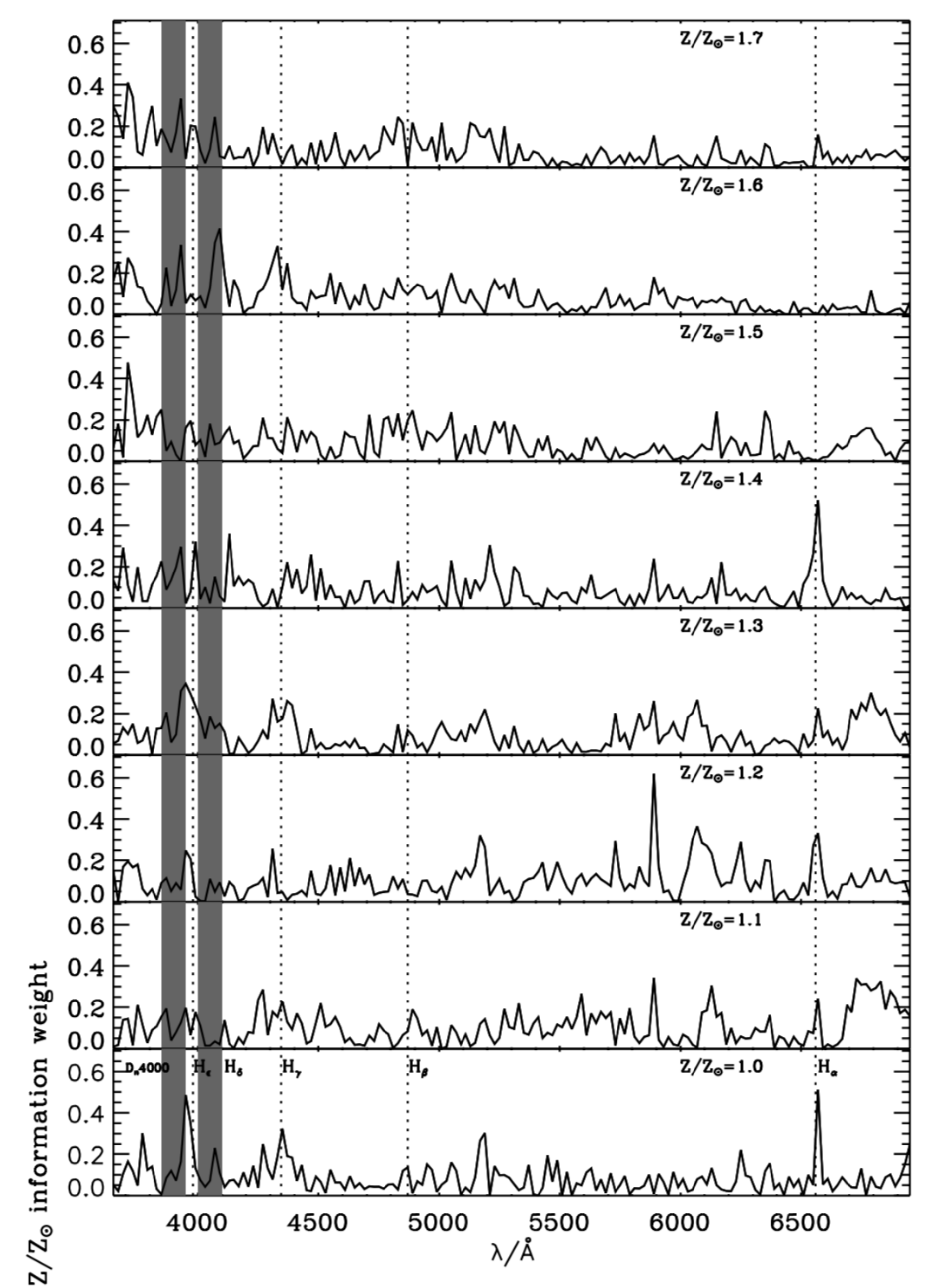}
  \hfill
  \caption{MOPED weights which show the information content as a function of wavelength for $8$ burst at ages $0.03, 0.23, 0.42, 0.66, 0.92, 2.56, 6.33$ and $14$ Gyr from top to bottom. The value of the weight (y-axis in arbitrary units) gives a measure of how sensitive the integrated spectrum is to the stellar component of a given age (left panel) and metallicity (right panel); these 16 variables are fitted simultaneously without any restriction.} 
     \label{fig:MOPED}
\end{figure*}

Noticeably, the young component has an impact not only on the shape of the 
continuum, affecting the $D4000$, but also on many lines that characterize 
the spectrum of the passive galaxy, as we will discuss in detail in Sect. 
\ref{sec:HK} and as clearly shown in Fig.~\ref{fig:MOPED}.
Medium-to-high resolution spectroscopy will then provide crucial information 
to discriminate this effect.

Finally, to complete our analysis and study the impact of different 
assumed values of $\alpha$-enhancement on our results, we also studied the newest 
\citep{vazdekis2016} models, considering BaSTI isochrones, different metallicities 
and values of $\rm[\alpha/Fe]$. 

We stress here that the aim of this paper is to address the impact of a possible contamination due to a young component on $H(z)$ measurements, while we defer to paper II a more complete discussion on how the assumed SPS models affects these measurements. However, we also note that this dependence has been also extensively studied in previous works. In particular, in \citet[][see Sects. 3.3 and 4 and Fig. 6]{moresco2012} and \citet[][see Sects. 4.2 and 5.1, Figs. 4, 5]{moresco2016} $H(z)$ measurements have been provided assuming two completely different SPS models to calibrate the method, obtaining measurements compatible at the 1$\sigma$ level, while in \citet[][see Sect. 4.3]{moresco2012b} and \citet[][see App. A]{moresco2016b} the impact of this effect has been studied on the estimated cosmological parameters, confirming that also in this case all the differences are well below the 1$\sigma$ level.


\section{Multiple stellar populations, where is the signal?}
\label{sec:youngcomp}

We begin by highlighting how different stellar populations of different ages 
contribute to the integrated stellar spectrum of a galaxy. To do this, we 
resort to the data compression algorithm MOPED \citep{reichardt2001}, which 
can massively reduce the number $p$ of data points, from $p$ to $m$, where $m$ 
is the number of parameters in the model, without losing information. In this 
process and when applied to a galaxy spectrum, MOPED computes the weights to be 
applied to the original data points for compression. By choosing as parameters 
of the model the amplitude of each of the single bursts of star formation that 
comprise the building blocks of a star formation history, we can visualize 
how each age contributes to the spectrum as a function of wavelength. This 
tells us where the signal for each age is located as a function of wavelength. 
These weights constitute eigenvectors: by construction they are orthonormal. 
Indicating by {\bf $x$} the input (data) spectrum (of length $p$) and by ${\bf 
b}_i$, the eigenvector for parameter (age bin) $i$, the weighted (compressed) 
data $y_i={\bf b}_i^T{\bf x}$ are uncorrelated. This is a very useful property 
as the information about each age ``bin" is uncorrelated (and uncorrelated with metallicity) . 

First, for the typical SDSS and VIMOS spectra (because of resolution, signal 
to noise and wavelength coverage) considered by 
\citet{moresco2012,moresco2015,moresco2016} it is only possible to 
reconstruct 8 age bins in the star formation history of a galaxy 
\citep{tojeiro2007}. We chose age bins such that the error in the recovered 
star formation history is minimized. The optimal bins are centered at 
$0.03$, $0.23$, $0.42$, $0.66$, $0.92$, $2.56$, $6.33$ and $14$ Gyr, which 
are very close to being equally spaced in log space. To illustrate where the 
information for age and metallicity are in the spectrum we construct a 
composite spectrum with constant star formation history (as to make all age 
components significant) and with increasing metallicity as to make this a free 
variable as well. All these $16$ variables are then recovered independently. 
The precise choice of the star formation and metallicity history will not 
change significantly the location of the information weights. The spectra are 
constructed without noise, 
since noise will be observation-dependent. The absolute value of the MOPED 
eigenvectors (i.e., the ``information weight", where a value above zero shows there is 
information and zero indicates lack of information) are shown in Fig. \ref{fig:MOPED} 
where ages are youngest to oldest from top to bottom.

There are several interesting features. First, for all ages, the age 
information is spread over the rest-frame visible wavelength range. Second, 
as expected, for the youngest stellar populations, the continuum carries 
significant information, but the absorption lines do also show significant 
weight. In particular, the most prominent features are the $D_n4000$ lines and 
$H_{\beta, \delta, \gamma, \epsilon}$. Note that for older stellar populations 
there is (localized) information all across the wavelength range. 

Finally, we can see that the youngest population is the most difficult to 
discriminate because most of the signal is in the continuum, but that 
information in the $D_n4000$, and in particular in \ion{Ca}{2} lines, remains. 
With appropriate S/N and moderate resolution spectra like those of SDSS and 
VIMOS (spectral resolution $R\gtrsim600$) it is possible to discriminate the 
young stellar component in passively evolving galaxies.

For the metallicity, the information is spread all over the absorption features 
and, very importantly, not all coincident with the age weights (by construction 
the eigenvectors are orthogonal); this shows that, with sufficient 
signal-to-noise, the age-metallicity degeneracy can be broken. 

This result is confirmed by many independent 
analyses. In \citet{connolly1995}, galaxy spectra were studied 
by orthogonal basis functions, providing the significant 
spectral components characterizing each particular galaxy type. 
\citet{chen2012} \citep[and later on][]{marchetti2013} used 
Principal Component Analysis to both classify and measure 
phisical properties of galaxies in the range $0.1<z<1$. 
Finally, \citet{yip2014}, from the analysis of SDSS 
simulated spectra, identified the region around the 
4000~\AA~break and H$\delta$ as primarily related to the age of 
the stellar population, with 
significant information contained also in the anti-correlation 
of the \ion{Ca}{2} H and K lines.

\subsection{\ion{Ca}{2} H/K diagnostic}
\label{sec:HK}

\begin{figure}[t!]
  \centering
  \includegraphics[width=\linewidth]{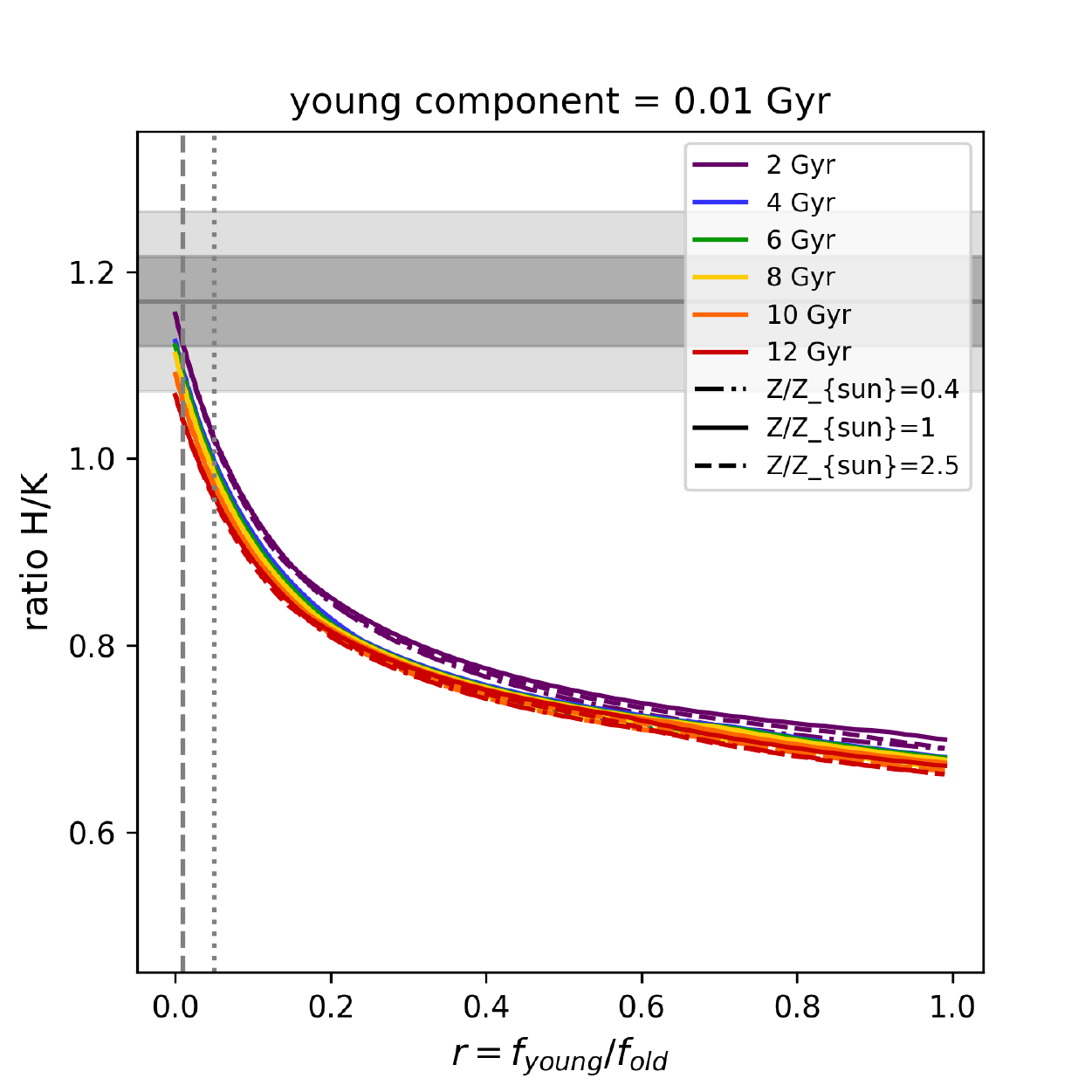} 
  \caption{Ratio between the \ion{Ca}{2} H and K lines as a function of the fraction between young (0.01 Gyr) and old component $r$. The dotted vertical line shows the value $r$ at which the two lines are equally deep ($r\sim0.05$), while the dashed vertical line the value of $r$ constrained by the data ($r<0.01$). The gray shaded area represents the range of $H/K$ values in the data (darker at 1$\sigma$, lighter at 2$\sigma$) from \citet{moresco2012,moresco2016}.}
     \label{fig:HK}
\end{figure}

\begin{figure}[t!]
  \centering
   \includegraphics[width=\linewidth]{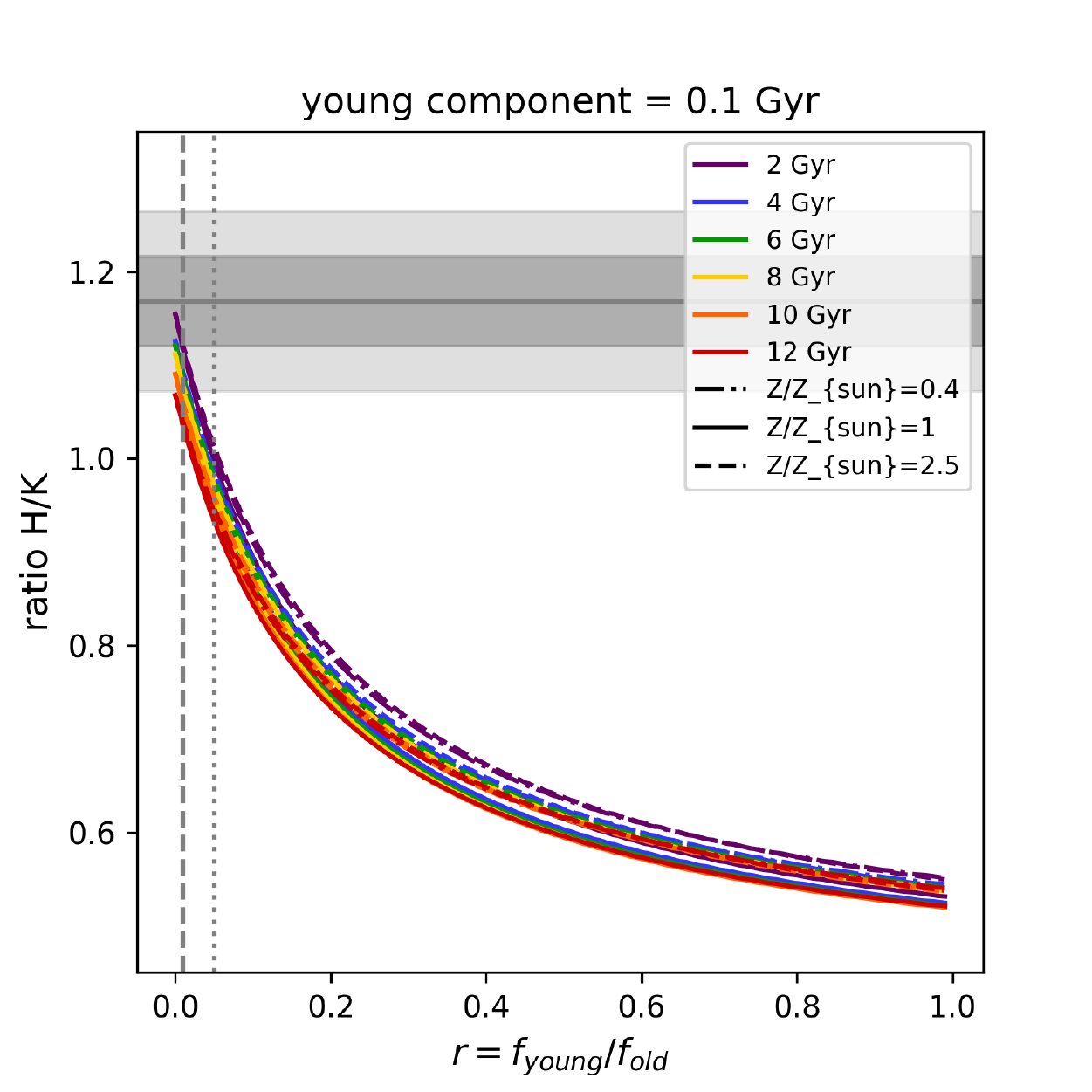}
     \includegraphics[width=\linewidth]{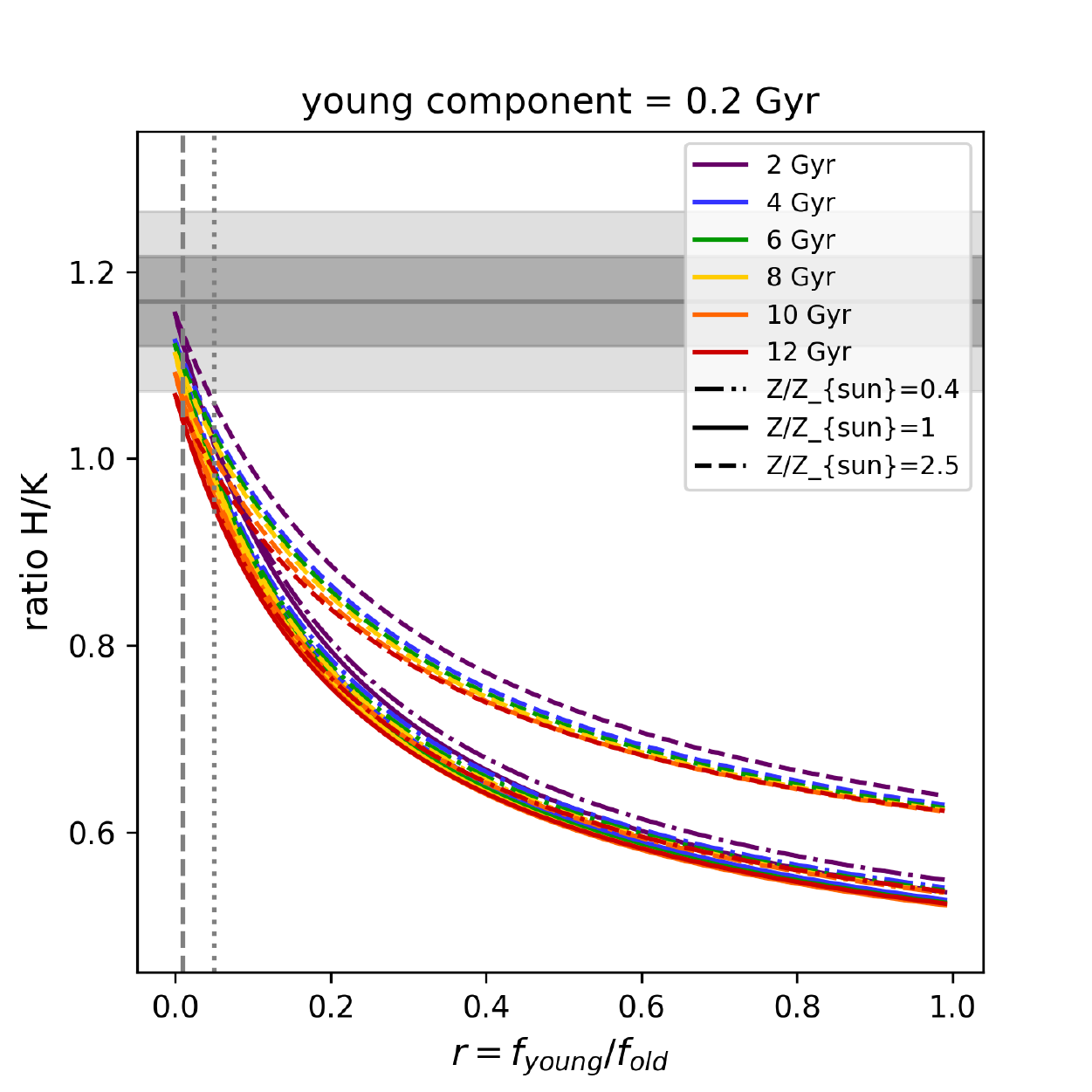}\hfill
  \caption{Same as Fig. \ref{fig:HK}. In this case the young component is 0.1 and 0.2 Gyr old.}
     \label{fig:HK1}
\end{figure}

Two of the main prominent features of the spectrum of a galaxy are the 
\ion{Ca}{2} K line at 3934~\AA~and \ion{Ca}{2} H line at 
3969~\AA, that partly define the $D4000$ too. 

In particular, in a galaxy dominated by a passive population it is 
always found that the K line is deeper than the H line, as can be 
seen in Fig. \ref{fig:spectrum}. On the contrary, this relation is 
{\em inverted} when a contribution of younger populations is present,
due in particular to the presence of H$\epsilon\lambda$3970~\AA~
that overlaps and gets combined with the \ion{Ca}{2} H line at 3969~\AA~(see Fig.
\ref{fig:spectrum}). Therefore, the ratio between the relative 
intensities of \ion{Ca}{2} H and K lines can be considered as an indicator of 
the relative contribution of a younger population to an older one. 

We define here the ratio $H/K$ as the ratio of the dip of
\ion{Ca}{2} H to the dip of \ion{Ca}{2} K, 
and we use this quantity to analyze the impact of contamination of a 
younger population.
This indicator was firstly proposed 
by \citet{rose1984} to constrain the ages of starbursts in post-starburst 
galaxies, and later on also adopted in different works 
\citep[see also][]{rose1985,leonardi1996,wild2007,sanchez2012}.

The metallicity of massive and passive galaxies are well constrained to be 
almost solar, or slightly over-solar, both in the local Universe \citep[see 
e.g.,][]{gallazzi2005,citro2016} up to $z\sim0.7$ \citep[see e.g.][]{moresco2016}, 
with some indications that this holds in general also at higher redshifts 
\citep[$z\sim1.6$, see][]{onodera2015}. Therefore, for the old component, we 
assume a solar metallicity. However, this is only done for convenience and to 
simplify the presentation of the results, as the metallicity of CC can be 
estimated from the spectrum as was done in \citet{moresco2016} (see also 
Fig.~\ref{fig:MOPED}). 
For the younger components, we explore three different ages, 0.01, 0.1, 
and 0.2 Gyr\footnote{For young components beyond 1 Gyr, the $H/K$ ratio is not 
inverted, but these components can be discriminated with other indicators, see 
Sect. \ref{sec:balmerlines}}, considering in each case three different 
metallicities, $Z/Z_{\odot}$=0.4, 1 and 2.5 for BC16 and 
$Z/Z_{\odot}$=0.5, 1 and 2 for M11. We consider a grid of values of $r$ between 0 
(when the population is totally dominated by the old component) to 1 (where the 
two population have the same weight). 

The results are shown in Figs. \ref{fig:HK} and \ref{fig:HK1}. We find that almost
independently of all the parameters (the age of the oldest 
population, the age of the younger contribution, its metallicity, 
the SPS model assumed), already at $r\gtrsim0.05$ there is an inversion of the 
$H/K$ ratio, with the \ion{Ca}{2} H line becoming deeper than \ion{Ca}{2} K. The 
figure shows also the \ion{Ca}{2} $H/K$ values obtained in the spectra analyzed in 
\citet{moresco2012} and \citet{moresco2016}, having a mean value
$<H/K>=1.17\pm0.05$ (at 1$\sigma$, $\pm0.1$ at 2$\sigma$). 
As can be seen by the figure, almost independently from the age
of the young component, these measurements are compatible with extremely low
values of $r$, namely $r\lesssim0.011$ (at 1$\sigma$, $r\lesssim0.027$ at 2$\sigma$). 
However, we emphasize that, in future analysis, the selection can be made even more 
stringent since the optical spectrum provides enough information to reduce $r$ below 
the $0.01$ level. 
Moreover, as we demonstrate in next sections, additional information can be 
gathered from other absorption and emission features in the optical spectra. 

Repeating the analysis procedure also considering the \citet{vazdekis2016} models 
with different $\rm[\alpha/Fe]$ ratios, we find results in complete agreement with 
the ones of Figs. \ref{fig:HK} and \ref{fig:HK1}, with a \ion{Ca}{2} $H/K$ inversion found at $r=0.037\pm0.003$ with varying ages and $\rm[\alpha/Fe]$ ratios, and with even more stringent constraints of the contamination of current measurements, with $r<0.011$ at 2$\sigma$. 
This demonstrates the robustness of this indicator against possible different levels of $\alpha$-enhancement.


\subsection{Presence (or absence) of [OII] and \texorpdfstring{H$\alpha$}{Lg} emission lines}
\label{sec:emlines}
The presence of an underlying young population has an impact also on emission 
lines in the optical wavelengths. \citet{magris2003} provided 
the theoretical expectations for equivalent widths (EW) of many emission
lines as a function of the age of the population and of the star-
formation history (SFH). If we focus in particular on the youngest ages 
($<$100 Myr), we find that almost independently of the assumed SFH,
strong equivalent widths are expected for both H$\alpha$ and 
[OII]$\lambda$3727, with values EW(H$\alpha)\gtrsim100-200$~\AA~ and
EW([OII])$\gtrsim40$~\AA.

These estimates assume quite prolonged SFHs, so that at the younger ages the 
population can be considered as star-forming, and dominated 
in the UV by O and B stars. We also explore the possibility of an 
instantaneous burst of star formation. 
We create photo-ionization models using the CLOUDY photo-ionization code
\citep[Version 13.03,][]{ferland1998,ferland2013}, assuming our galaxies as 
formed by a central ionizing source surrounded by a spherical cloud.
We simulate the central ionizing source using the same models discussed
in Sect. \ref{sec:method}, where a young component (with an age of 0.01, 0.1 or 0.2 Gyr)
is superimposed on an older component (with an age of 4, 8 or 12 Gyr) with different 
fractions $r$; this allows us to extrapolate the results as a function of the various
parameters considered in our analysis. We assume a starting ionization parameter 
log(U)=-3.2, which is consistent with the observations of local \ion{H}{2} 
regions \citep[$\rm -3.2 <log(U)< -2.9$, see][]{dopita2000} and star-forming 
galaxies \citep[see][]{moustakas2006,moustakas2010}. 
For the ionized nebula, we assume a fixed hydrogen density $\rm n_{H} =100~ 
\rm cm^{-3}$, which is the typical density of observed star-forming regions 
\citep{dopita2000,kewley2001,dopita2006}.
We adopt the solar chemical composition by \citet{asplund2005}, matching the 
gas metallicity of the ionized nebula with the stellar metallicity of the 
ionizing population. Moreover, we account for the presence of dust adopting 
the default CLOUDY ISM grains distribution. 

We find that both [OII] and H$\alpha$ emission lines are extremely sensitive to
the presence of the youngest and hottest stars with ages $\lesssim$10 Myr,
with significant fluxes already with a contamination $r=0.05$, and in particular $F(H\alpha)\sim0.2-0.9\times10^{-15}$ erg/s/cm$^{-2}$ and $F([OII])\sim0.5-3\times10^{-15}$ erg/s/cm$^{-2}$ (depending on the age of the old component); similarly, at the same level of contamination we also find relatively high values of equivalent widths, $EW(H\alpha)\sim3.5-9.5$ and $EW([OII])\sim7-30$. These values increase by $\sim$70\%
when the contamination is $r\sim0.1$.
At older ages of the young component ($\gtrsim100$ Myr), there is a significant drop of UV emission \citep[as can be seen also in][]{levesque2010,citro2017}, so that it is no more able to produce significant emission lines; however, these other phases can be discriminated with other indicators (see Sect. \ref{sec:HK}).



To select optimal probes for the cosmic chronometers approach, and 
minimize the possible contamination due to a younger population, it is 
therefore crucial to carefully select galaxies not only on the basis of  the photometry, but also from spectroscopy. In 
\citet{mignoli2009}, a criterion was proposed to separate passive and 
star-forming galaxies on the basis of a cut in [OII] ($\rm EW([OII]<5$~\AA).
Similarly, in \citet{wang2018} passive galaxies were separated from the other
populations on the basis of the detectability of H$\alpha$ emission line,
and in particular having $\rm S/N_{H\alpha}<3$.
In \citet{moresco2012,moresco2016} galaxies have been carefully 
selected by rejecting objects with detectable emission lines in H$\alpha$ 
and [OII], with EW thresholds that are compatible, given the previously 
discussed models, with no UV emission.
In particular, in \citet{moresco2012} a criterion EW([OII])$<$5~\AA~ was applied
\citep[according to][]{mignoli2009}, and in \citet{moresco2016} 
EW([OII])$<$5~\AA~ and a S/N([OII], H$\alpha$, H$\beta$, [OIII])$<$2. The 
analysis of the stacked spectra highlighted in both cases no evidence for 
emission lines.

\begin{figure*}[t!]
  \centering
  \includegraphics[width=0.9\linewidth]{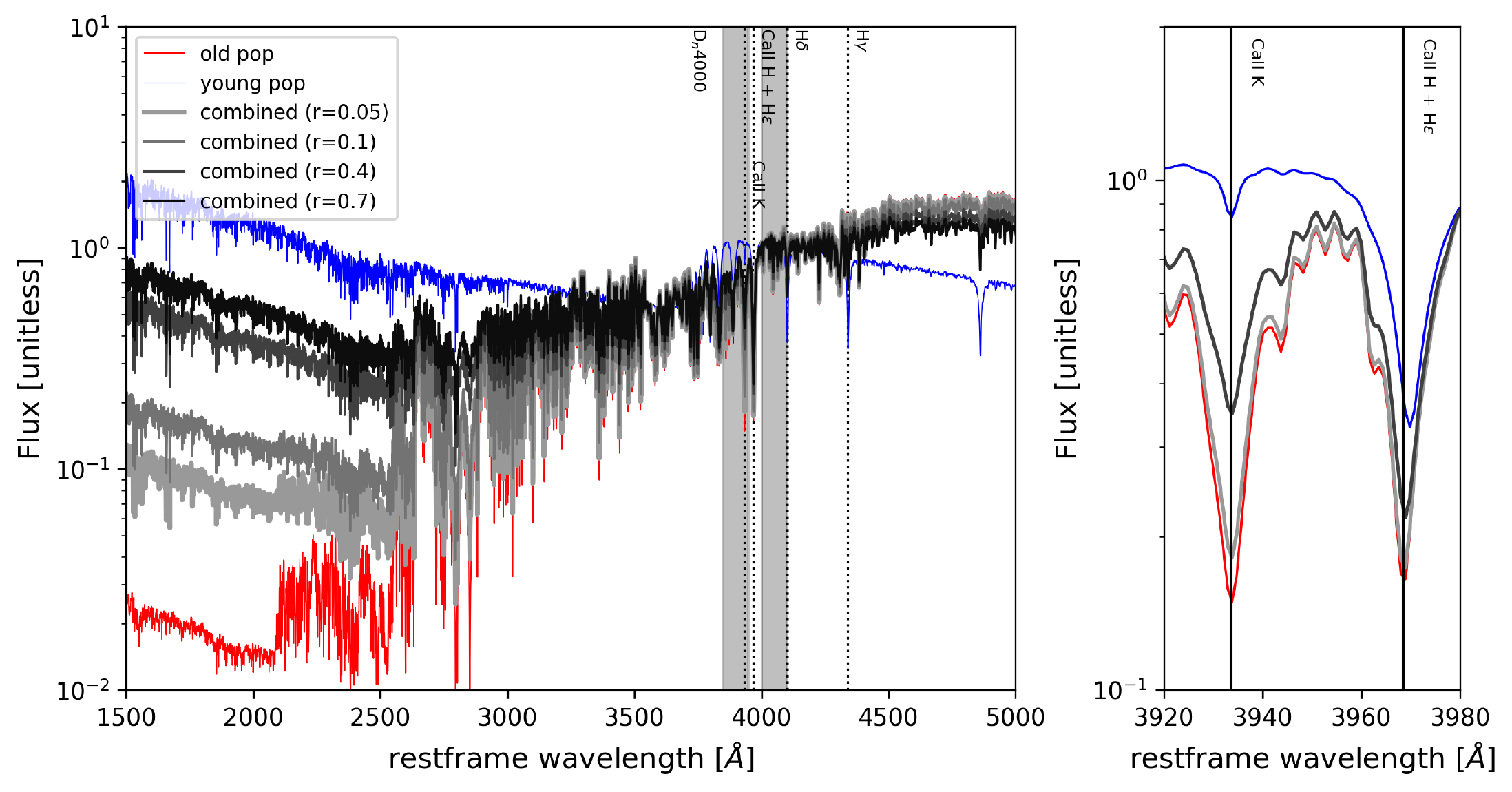} 
  \caption{Contribution at UV wavelengths of different percentages of a young stellar component. The red and blue curves represent an old (8 Gyr) and a young (0.1 Gyr) component, combined with different ratios to produce the black and gray curves ($r=0.05, 0.1, 0.4, 0.7$). The impact in the UV can be seen by the total flux produced between 1500 and 2500~\AA. The right panel shows a zoom-in around the \ion{Ca}{2} H and K lines, showing its inversion as a function of $r$.}
     \label{fig:UV}
\end{figure*}

\subsection{The UV contribution}
\label{sec:UV}
Another significant tracer of star formation is the UV flux 
\citep{kennicutt1998}. There are many observational evidences that some 
elliptical galaxies show the presence of UV emission. The mechanism behind this 
emission is still debated, with suggestions that there could be other processes 
than the presence of a young component producing an UV excess \citep[e.g., see][]{greggio1990};
therefore, any conclusion about a young component based on the presence of UV 
emission should be carefully considered as an upper limit. Typically, these
processes can be discriminated on the basis of some color-color diagrams. In particular, 
\citet{yi2011} \citep[but see also][]{han2007,donahue2010,boissier2018} proposed a 
combination of UV and optical colors to separate the UV emission coming from a young 
component or a UV excess. They suggested that a combination of $FUV-NUV$ and $NUV-r$ 
colors allows to separate these components, with $FUV-NUV>0.9$ indicating a negative UV 
slope (hence showing evidences for an UV emission), and $NUV-r<5.4$ indicating an 
emission due to a young component (and respectively $NUV-r>5.4$ an emission due to an 
UV excess).
Recently, \citet{ilbert2013} proposed another selection criterion based
on UV colors, the $NUVrJ$ diagram, that was found to be particularly sensitive
to young ages \citep[0.1-1 Gyr,][]{martin2007,arnouts2007}. This 
selection criterion was successfully applied by many authors to safely discriminate 
between star-forming and quiescent populations 
\citep{ilbert2013,ownsworth2016,moutard2016,davidzon2017}, as opposed to the 
other criteria \citep[$UVJ$ and $NUVrK$, respectively][]{williams2009, 
arnouts2013}, and can be therefore used, in combination to the previously 
discussed ones, to provide a pure, passively evolving sample of CCs.
The advantage of this criterion, given its large wavelength coverage, is that it 
is also useful to disentangle effects due to age and extinction, and in particular
to discriminate possible UV emission hidden by the presence of dust.
To explore how much a young component could contribute to UV fluxes (in this case 
neglecting the possible contribution of obscuration due to dust), in Fig. 
\ref{fig:UV} we compare the combined spectra of an old (8 Gyr) and a younger 
(0.1 Gyr) component built with different ratios $r$. 
We find that small value of $r$ produce a 
flattening in the spectrum at UV wavelengths, but with fluxes $\sim$1 order of 
magnitude smaller than in the optical, while at relatively high values of 
$r\gtrsim0.4$ we obtain a more significant excess UV flux. As demonstrated in 
Sect. \ref{sec:HK}, these values of $r$ can be discriminated on the basis of the 
$H/K$ diagnostic, as also shown in Fig. \ref{fig:UV}. Therefore, the $H/K$ and 
the UV flux diagnostic are highly complementary, and can be used in combination to 
improve the the purity of the CC sample selection.


\subsection{Higher-order Balmer lines}
\label{sec:balmerlines}
Higher order Balmer lines, such as H$\delta$ and H$\gamma$, are also powerful 
tools to discriminate recent episodes of star formation. {} This is easy to 
understand since these lines are due to the abundance of H, which changes very 
little between different values of the metallicity content of galaxies (about 
1\%), therefore it measures, effectively, the gravity of the stars, thus their 
age. In particular, strong H$\delta$ absorption lines (EW(H$\delta)>4-5$~\AA) have 
been historically used to identify post-starburst (or H$\delta$ strong) galaxies, objects that have ceased to form stars in the 
past (EW([OII])$>-2.5$~\AA) but experienced recent episodes of star-formation 
\citep[e.g. see][]{leborgne2006,wilkinson2017}.
These indicators have been used to trace this population up to $z\sim1.5$ \citep{bezanson2013}. Differently from H$\epsilon$, these lines do not overlap with other absorption lines, producing a well recognizable pattern as the one described for the \ion{Ca}{2} $H/K$ inversion in Sect. \ref{sec:HK}; hence, they can be used to trace star-formation on slightly longer time-scales than the $H/K$ diagnostic \citep[typically between hundreds of Myr up to $\sim$1-2 Gyr, as discussed in][]{leborgne2006,bezanson2013,wilkinson2017}, and given the age of the young population, they are associated to lower values of UV emissions.
Therefore, combining this indicator with the $H/K$ ratio it is possible to 
cover a wide range of recent episodes of star formation in the galaxies.

\begin{figure*}[t!]
  \centering
  \includegraphics[width=0.49\linewidth]{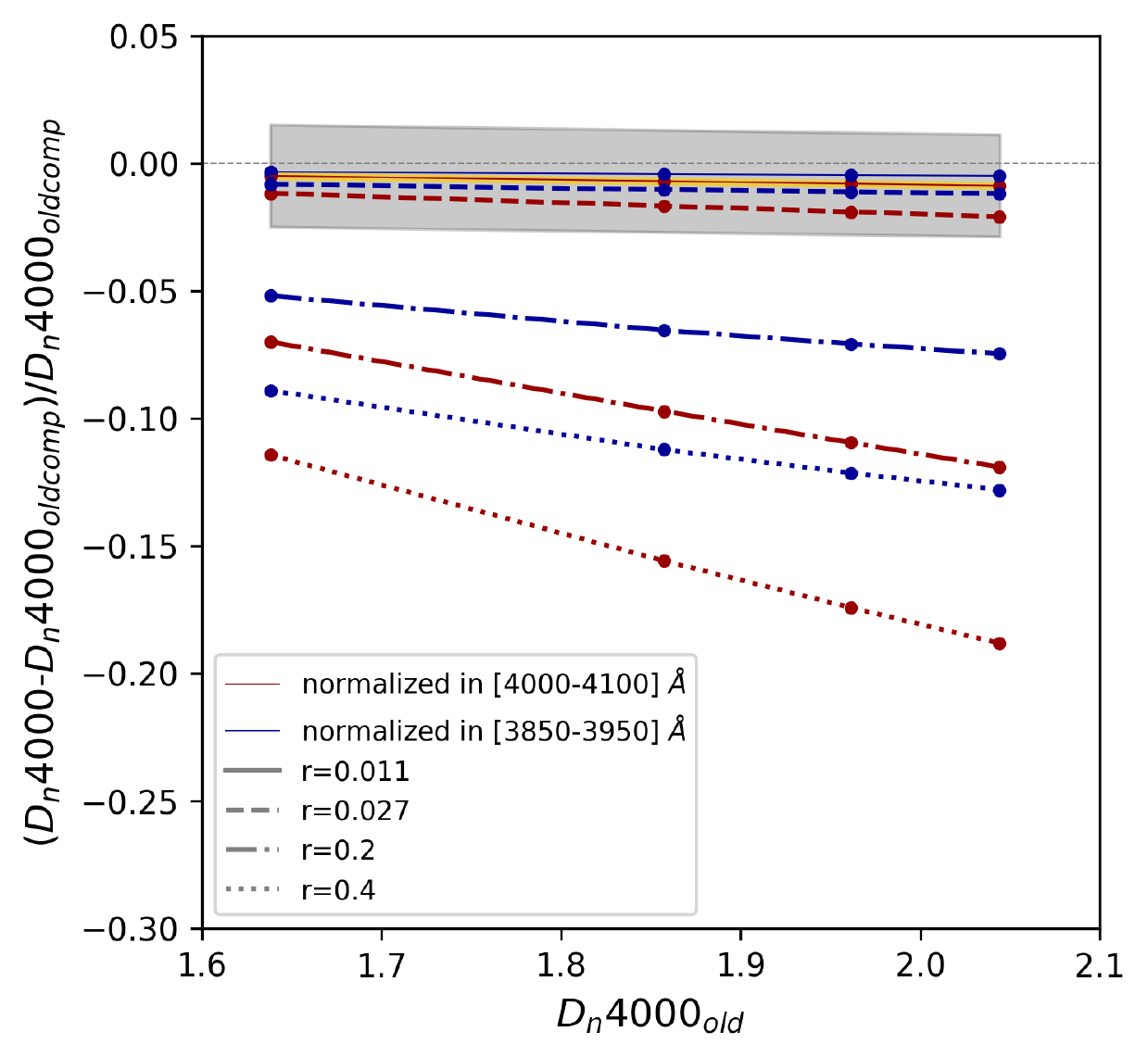}
  \includegraphics[width=0.49\linewidth]{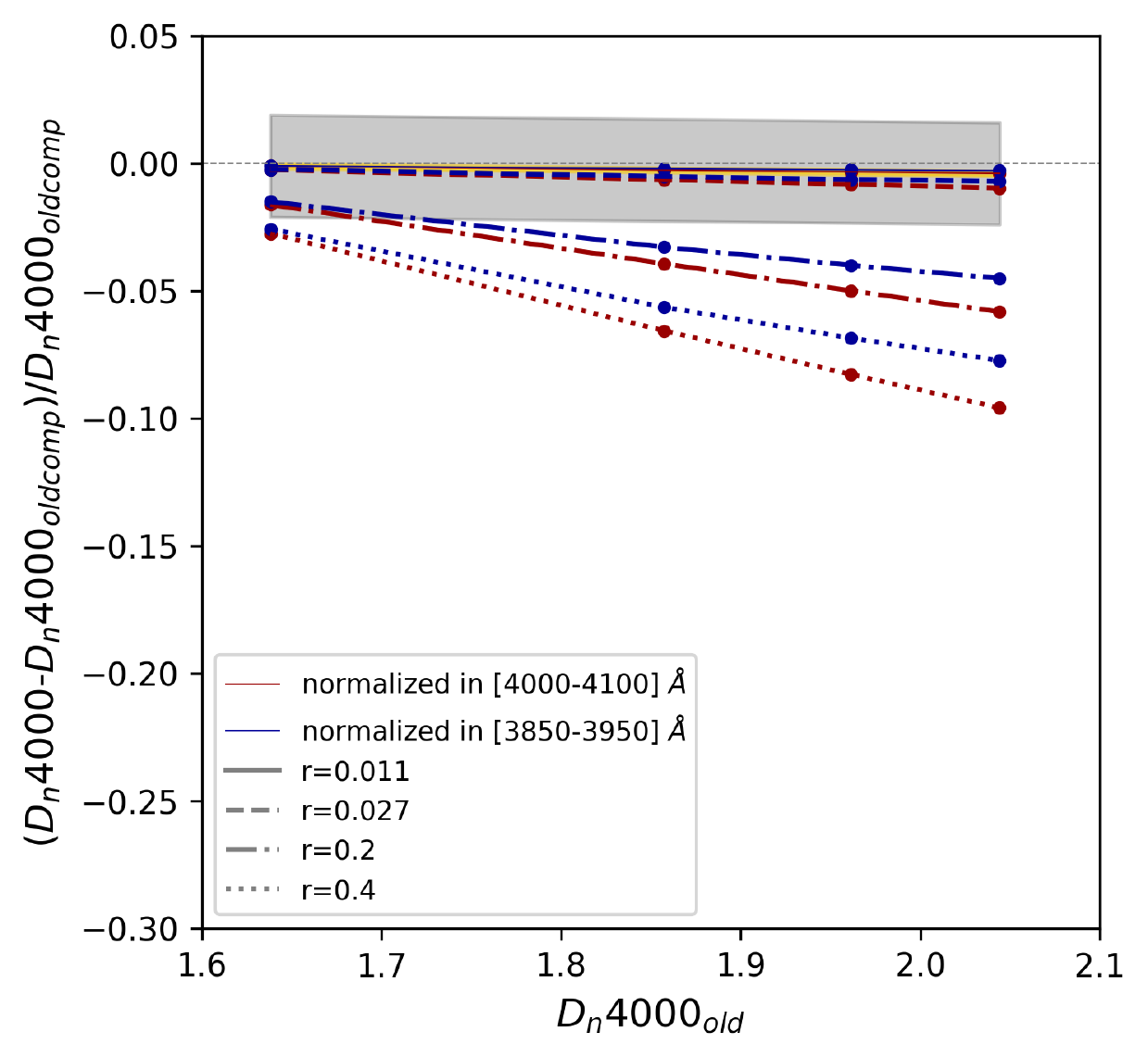}
   \hfill
  \caption{Percentage variation of $D_n4000$ due to the presence of a young component, for different values of $r$; in the left panel is shown the case of a 0.1 Gyr young component, in the right panel a 1 Gyr young component. We explored two different combinations, with young and old components fluxes normalized in the range [4000-4100]~\AA~($r$, red lines), and in the range [3850-3950]~\AA~($r'$, blue lines). Solid, dashed, dot-dashed and dotted lines correspond respectively to values of $r=0.01,0.1,0.2,0.4,0.6$. The shaded area represent a 2\% (grey) and 0.1\% (yellow) statistical error on $<D_{n}4000>$, which is the typical error found in \citet{moresco2012,moresco2016} that can be achieved with a larger ($N_{gal}\gtrsim10000$) and smaller statistical sample ($N_{gal}\lesssim1000$), respectively. 
  }
     \label{fig:D4000}
\end{figure*}

\vspace{5mm}
\section{The impact of a young component on the \texorpdfstring{$D_n4000$}{Lg} and Hubble parameter }
\label{sec:D4000}
The $D_n4000$ break is a discontinuity in the spectrum of galaxies at 4000~\AA~
rest-frame due to the blending of several absorption features, and 
defined as the ratio of the mean flux $<F_{\nu}>$ between a red and 
a blue band-pass:
\begin{equation}
D_{n}4000=\frac{(\lambda_{2}^{blue}-\lambda_{1}^{blue})\int_{\lambda_{1}^{red}}^{\lambda_{2}^{red}}F_{\nu}d\lambda}{(\lambda_{2}^{blue}-\lambda_{1}^{red})\int_{\lambda_{1}^{red}}^{\lambda_{2}^{blue}}F_{\nu}d\lambda}
\label{eq:D4000}
\end{equation}
There have been different definitions of these bands, in
particular a broader definition with [3750-3950], [4050-4250]~\AA~
\citep{bruzual1983,hamilton1985} and a narrower definition with
[3850-3950], [4000-4100]~\AA~ \citep{balogh1999} which is less
sensitive to reddening effects; here we adopt the narrow definition.
It is important to bear in mind that $D_n4000$ has been originally defined in flux $F_{\nu}$, and not in $F_\lambda$.

This index of known to depend on the age and metallicity of the 
stellar population \citep[e.g., see][and the results obtained in Sect. 
\ref{sec:youngcomp}]{poggianti1997}. For this reason first in \citet{moresco2012}, 
and later on in \citet{moresco2015} and \citet{moresco2016}, the corresponding break 
was used to relate explicitly Eq. \ref{eq:Hz} with Eq. \ref{eq:Hz_Dn}, and then was 
used to obtain a measurement of $H(z)$ from the spectral evolution of 
CC. It was argued that this choice of (non-lossless) data compression is useful to 
decouple systematics and statistical effects in the estimate of the age of the 
stellar population.
It has been also verified that $\alpha$-enhancement has a negligible impact on this 
spectral feature \citep[][see App. A.3]{moresco2012}, representing this another 
advantage for moving into D4000-z space, instead of age-z space.

Here we test if, and quantify how much, this index may be affected by the 
presence of an underlying younger stellar population.

The impact of a young component on the $D_{n}4000$ is quantified
in Fig. \ref{fig:D4000}, where we show the percentage variation of 
$D_{n}4000$ when including a young component for different values of $r$ 
and $r'$. To explore the dependence of the results on the age of the young 
component, we consider two different values, 0.1 and 1 Gyr, that, as discussed 
in Sect. \ref{sec:HK} and \ref{sec:balmerlines}, have a different impact on the 
composite spectrum.
 
As it can be seen, the deviation from the unbiased measurement, which would 
have an impact also on cosmological inference, is more and more significant 
at increasing values of $r$. In the case of the youngest component (0.1 Gyr), 
the effect on $D_n4000$ ranges from $\lesssim$5\% when $r<0.05$, 
up to $\gtrsim$20\% when $r>0.4$ (we note that, as discussed in Sec. \ref{sec:HK}, 
at $r=0.05$ we are able to distinguish a young component from the $H/K$ 
inversion). In the case of the oldest component (1 Gyr) instead, the impact is 
always $\lesssim$10\%.
At the same time, we see that in both cases the impact can be kept at a negligible level, at
least for current data, if the contribution is limited to 
small values ($r\lesssim0.01$). 

Recall that since the CC approach is based on a {\em differential} measurement
of ages (or $D_n4000$), what would actually impact the cosmological constraints
is not a systematic offset (in that case, only the absolute age or $D_n4000$ 
measurement would be biased), but rather the relative difference between 
different $D_n4000$ measurements, i.e., the slope of the relations shown in Fig. 
\ref{fig:D4000}. The statistical error associated to the 
measurement of the mean $\langle D_{n}4000\rangle $ can be estimated from current 
data \citep{moresco2012,moresco2016}; we find that on average the statistical 
percentage error ranges between 2\% \citep[in regimes with a smaller statistic, 
with $\lesssim 1000$ objects, corresponding to the $z>0.5$ analysis][]
{moresco2012} and 0.1\% \citep[in SDSS-like regimes, with $\gtrsim 
10000$ objects][]{moresco2012,moresco2016}. In fig. \ref{fig:D4000} only the low 
statistics case can be seen as the other one is almost coincident with the line 
width.

We find that with a lower number of galaxies, the capability of 
distinguishing a young component is reduced due to the larger error, and therefore 
its impact on cosmological measurements should be by definition sub-dominant in
the total error budget; with a larger statistics, it is instead fundamental to 
assess the percentage of contamination $r$.

\begin{figure}[t!]
  \centering
  \includegraphics[width=1.\columnwidth]{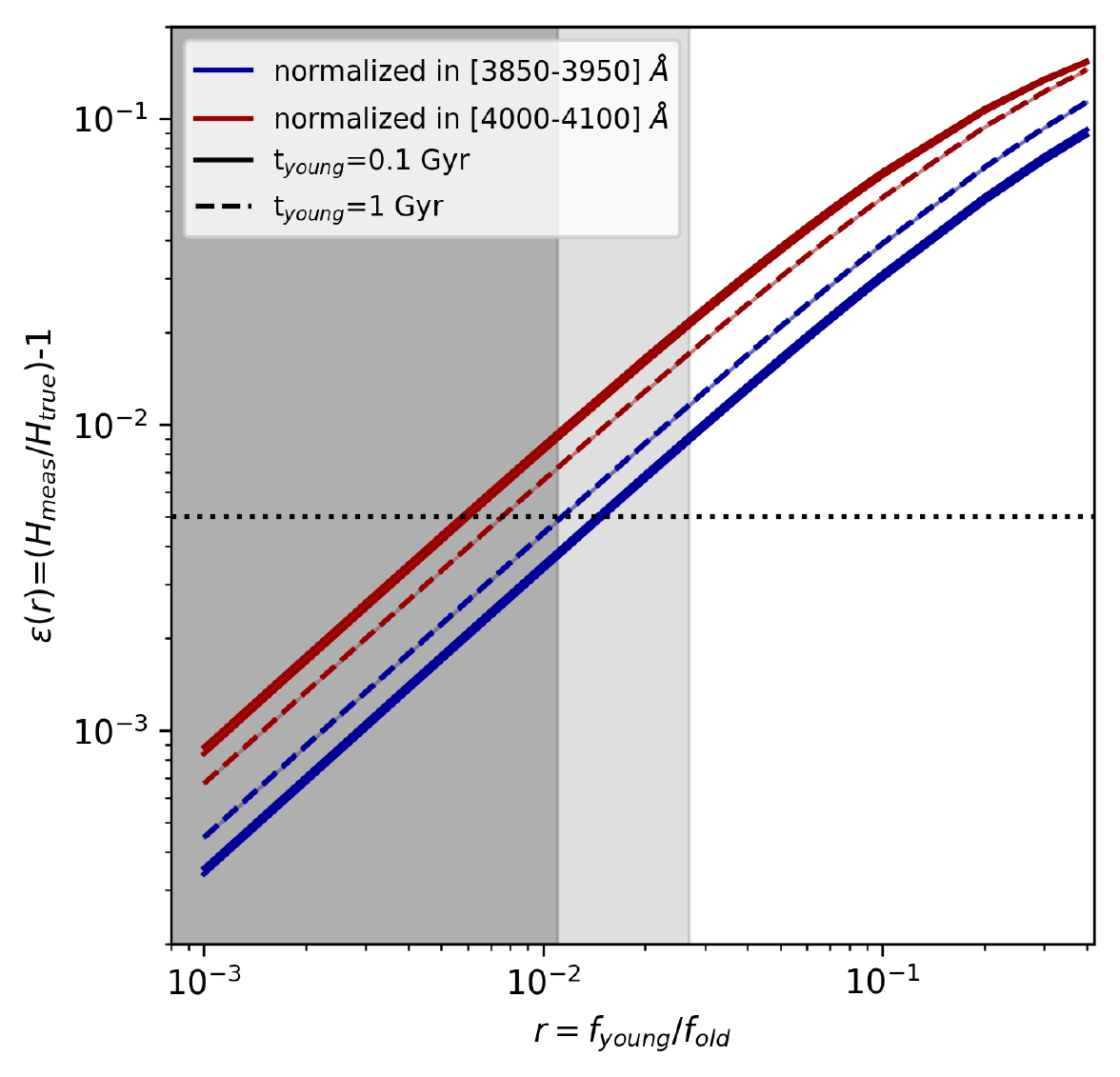} 
  \caption{Impact on the measure of $H(z)$ of a young component with fraction $r$ with respect to the old component. The different colors show the effect when the spectra are normalized in different wavelength ranges ([4000-4100]~\AA~for the red curves, [3850-3950]~\AA~for the blue curves), while the different line-styles represent different ages of the young component (0.1 and 1 Gyr). The grey shaded area show the ranges of $r$ allowed by the data in \citet{moresco2012,moresco2016} averaged across all redshifts (darker at 1$\sigma$, lighter at 2$\sigma$), and the black dotted line shows an error $\epsilon(r)$=0.5\%. The shaded area (barely visible) between the curves represent the uncertainty due to different SPS models.}
     \label{fig:Hz}
\end{figure}

The impact of the presence of a young component with a fraction $r$ on $H(z)$
is estimated in Fig. \ref{fig:Hz}. From Eq. \ref{eq:Hz_Dn}, we derive an analytic expression of the percentage deviation of $H(z)$ with 
respect to the true one:
\begin{equation}
\frac{H(z)_{meas}}{H(z)_{true}}=\frac{dD_n4000_{true}}{dD_n4000_{meas}}=1+\epsilon(r)\;,
\label{eq:Hz_err}
\end{equation}
where $\epsilon(r)$ is the percentage variation of $dD_n4000$ as a function of the 
young component fraction $r$, namely $(1-\alpha)^{-1}$ where $\alpha$ is 
the slope of the relations in Fig. \ref{fig:D4000}. 
We note that formally $r=r(z)$, since the contamination due to a young component could 
be different depending on the couple of points used to estimate 
H(z); the only assumption we make here is that $r$ is constant between the two 
redshifts considered to provide one $H(z)$ measurement \citep[that should, therefore, 
be really close in cosmic time to avoid evolution effects, as done in][]{moresco2012,moresco2016}.
The validity of this assumption can be directly tested on the data. We note here, however, 
that in CC approach one would more likely work not comparing individual galaxies (where this 
effect is actually maximized), but comparing mean trends of populations very close in cosmic 
time and selected homogeneously, helping to keep this effect under control.

In the figure we show $\epsilon(r)$ as a function of the age of the young component 
(0.1 and 1 Gyr) and of the chosen range of normalization. As can be seen, there is a 
large variation depending on the considered parameters, but we find that a young 
component effectively acts on $H(z)$ as an additive systematic bias, that can 
influence the measurement up to 20\% if not properly taken into account, i.e. for 
a contribution $r\sim0.4$. As shown, the impact due to the different 
assumed SPS model is, instead, negligible.
It is therefore crucial to limit this effect, as done in 
\citet{moresco2012} and \citet{moresco2016} by selecting an uncontaminated (or  
minimally contaminated) sample.

The $H/K$ values found in these works (shown in Figs. \ref{fig:HK} and \ref{fig:HK1})
allow for a 
maximum value of $r=0.009-0.013$ (at 1$\sigma$, depending on the age of the youngest 
population, $r=0.023-0.035$ at 2$\sigma$), which is, however, a strong upper limit 
on the contamination level of our sample, since the same value of $H/K$ can be 
reproduced by a purely passive population at a different age. 
This effect translates to a maximum variation 
$\Delta D_{n}4000<$0.6-1\%, which, as a consequence, affects the Hubble 
parameter at most by 0.4-1\% (at 1 $\sigma$, considering the full variation spanned by 
the various models shown in Fig. \ref{fig:Hz}, 0.8-2.3\% at 2$\sigma$). This is below 
current uncertainties on the derived values of $H(z)$ with this method. We also
estimate that for this systematic effect to be $<0.5$\%, the contamination by a young component 
should be $r\lesssim0.005-0.015$ (depending on the model).

In general, for unrelated errors the associated covariance matrix (Cov)
for the Hubble parameter $H(z)$ is simply the sum of the covariance 
due to the various errors:
\begin{equation}
\rm Cov_{ij}^{tot}=Cov_{ij}^{stat}+Cov_{ij}^{young}+Cov_{ij}^{model}+Cov_{ij}^{met}
\end{equation}
where ``stat", ``young", ``model" and ``met" denote the contributions to the covariance 
due to statistical errors, young component contamination, dependence on the chosen 
stellar population model, and metallicity respectively. These error contributions will 
be studied in details elsewhere. 
Here, we just provide the equation for $\rm Cov_{ij}^{young}$.
Following Eq. \ref{eq:Hz_err}, at redshift $z_i$ the systematic 
error on the measurement of the Hubble parameter due to a young 
component residual is $H_{true}\epsilon(r_i)$; therefore, its 
associated covariance matrix will be:
\begin{equation}
{\rm Cov}^{\rm young}_{ij}=H_{true}(z_1)H_{true}(z_2)\epsilon(r(z_1))\epsilon(r(z_2))\,.
\label{eq:covy}
\end{equation}
The error on $H$ arising from a possible young component contamination is highly 
correlated across redshifts. However, even in current analyses, it accounts for $\sim 
0.5-1\%$ correlated error, while the statistical error for each data point is 5\% or 
larger; hence it is a sub-dominant contribution. 

\section{Towards an optimal selection of cosmic chronometers}
\label{sec:optsel}
\begin{figure}[t!]
  \centering
  \includegraphics[width=1.15\columnwidth]{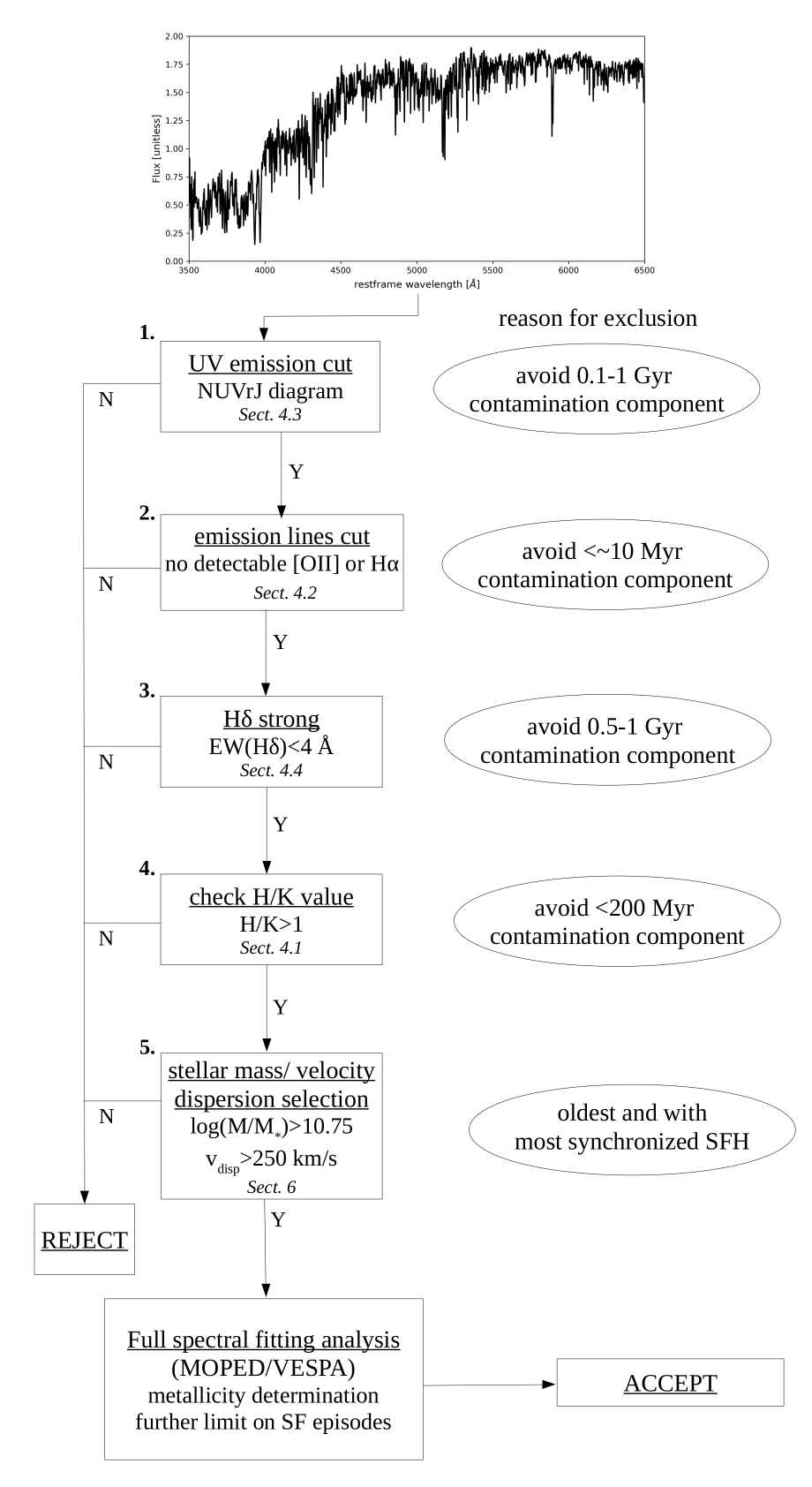} 
  \caption{Work-flow to optimally select cosmic chronometers. This is the procedure that we have applied in our previous works. Note that {\em all} boxes need to be checked to select a cosmic chronometer. Selection criteria can be made more stringent for future data with improved statistics.}
     \label{fig:flow}
\end{figure}

Finally, based on all the results discussed so far, we provide a recipe
to optimally select cosmic chronometers for a $D_n4000$ analysis, to minimize 
the possible effects due to an underlying young component, and, therefore, to 
maximize the robustness of the cosmological results.
As discussed in \citet{moresco2012,moresco2016}, the selection criterion
is crucial in the $D_n4000$ approach to cosmic chronometers to select the purest 
possible sample of massive and passively evolving galaxies. Many criteria have 
been combined based on both photometry and spectroscopy to obtain 
a pure sample of massive and passively evolving galaxies\footnote{In 
\citet{moresco2013} it is shown that this combination of criteria, while reducing 
the sample size by $\sim$70\% with respect to a simple color cut, it also significantly 
reduces the level of contamination.}. 
Here we summarize the main ones, discussing also the possible effect of not 
considering (possibly due to unavailability of data) one of those. 

The fundamental steps to select the best cosmic chronometers candidates are:
\begin{enumerate}
\item {\em Selection based on UV emission.} As discussed in Sect. 
\ref{sec:UV}, having UV coverage gives a first photometric indication if there 
could be a young component, which can not be visible and distinguishable in 
optical photometric bands. A selection based on UV colors, such as the NUVrJ 
color-color diagram \citep{ilbert2013}, can detect the presence (even at low 
level) of a young population from few hundreds of Myr up to $\sim$1 Gyr, 
and therefore galaxies not matching this criterion should be rejected.
\item {\em Check for the presence of emission lines.} As discussed in Sect. 
\ref{sec:emlines}, the presence of H$\alpha$ and [OII] emission lines is a 
spectroscopic indicator of on-going star-formation, due to the contribution of 
young stellar population with ages $\sim$10-100 Myr. While different criteria 
can be adopted to minimize this contamination, either based on EW values 
($\rm EW([OII])<$5~\AA, see \citet{mignoli2009}) or on the signal-to-noise ratio 
of these emission lines \citep[$\rm S/N_{H\alpha}<3$, see][]{wang2018}, 
only spectra with no detectable emission lines should be 
considered. It is fundamental to reject galaxies showing any evidence of 
emission lines.
\item {\em Check for the presence of strong H$\delta$ absorption line.} 
Complementary to the selection (2), a cut based on the presence of 
H$\delta$ absorption line \citep[EW(H$\delta$)$>$4-5~\AA,][]
{leborgne2006,bezanson2013,wilkinson2017} is fundamental to exclude 
galaxies that have experienced a recent episode of star formation, yielding a young stellar component of $\sim0.5-1$ Gyr (post-starburst galaxies). 
Galaxies with values above that threshold should be therefore excluded by the selection.
\item {\em Check for \ion{Ca}{2} $H/K$ value.} In Sect. \ref{sec:HK}, we 
demonstrated how the \ion{Ca}{2} $H/K$ ratio is a powerful tool to 
diagnose the presence of an underlying young component. A check of the 
\ion{Ca}{2} $H/K$ value found in the data should allow to assess
the corresponding level of contamination in the data, and therefore the
corresponding contribution to the total error induced on $H(z)$ following
the recipes provided. In particular, we show that galaxies that present an inversion 
of this indicator ($H/K<1$) are characterized by the presence of a young component 
(with ages $\lesssim$ 200 Myr) with a percentage $r\gtrsim$0.05, and that values of 
$H/K>1.1$ indicates a contamination $r\lesssim 0.01$.
\item {\em Stellar mass/velocity dispersion selection.} Finally, it is fundamental 
to select, amongst the previously selected passive galaxies, the most massive 
ones. In particular in \citet{thomas2010} it was shown not only that with 
increasing mass the formation of galaxies dates progressively back in the past 
(therefore ensuring selection of the oldest objects at each redshift), but also 
that their SFH becomes more and more synchronized, making them the ideal CC. 
Moreover, in \citet{moresco2013} it was also shown how the mass cut increases 
the purity of the sample, since contamination is more important at smaller 
masses. Both a cut in stellar mass and velocity dispersion can be 
applied to select the most massive objects, where the second one, being a direct 
observable, is less dependent on model assumptions. A typical cut in stellar mass 
adopted is $\rm log(M/M_{\odot})>10.75-11$, roughly corresponding to a cut in 
velocity dispersion $\rm v_{disp}>250-300$ km/s.
\end{enumerate}

The combination of the $H/K$ diagnostic combined with as many independent observable as 
possible allows to avoid statistical outliers in the selection, and minimize the 
contamination.
At last, the selected spectra should be analyzed with a full-spectral fitting 
approach like MOPED/VESPA or similar approaches 
\citep[e.g.][]{chevallard2016,citro2016}. 
This step represents a final cross-check, that will be applied on a sample that passes the 
selection criteria discussed above, to detect possible further signs of star formation that 
have not been detected by other observables. In this way, the selection will rely as little 
as possible on modeling and fitting, and as much as possible on observables, providing as a 
consequence an Hubble parameter measurement less model-dependent.
As discussed in Sect. \ref{sec:youngcomp}, with enough signal-to-noise in the data this 
kind of analysis is able not only to provide constraints on the metallicity of the 
galaxy, needed to apply the CC method, but also to disentangle the galaxy SFH, and 
identify the presence of further star formation episodes that were not excluded with 
the previous criteria \citep[similar results have been also obtained with independent 
approaches, see e.g.][]{citro2016}.
This approach maximizes the purity of the CC candidates, providing a suitable 
sample to constrain the expansion history of the Universe.

As discussed before, some information may be unavailable; however, there is 
some overlap between the criteria, in particular between criterion (1), (2) and (4). 
On the one side, the absence of \ion{Ca}{2} 
$H/K$ inversion limits the possible level of contribution of a young 
component also in the UV, on the other side the absence of significant 
emission lines is directly linked to the underlying absence of young stars. A 
combination of criteria (1,3,4,5) or 
(2,3,4,5) could be, therefore, sufficient to select CC. However redundancy offers a 
more reliable selection.
Selection criteria can (and must) be made most stringent for future data with increased 
statistics.

The selection limits discussed above are clearly dependent on resolving power and 
S/N of the spectrum. The most critical features to be measured are \ion{Ca}{2} 
$H$ and $K$ lines, and we find that, in order to detect them, medium- to low-resolving 
power is needed, and currently they have been detected down to $R\gtrsim230$ 
(with VIMOS low-resolution instrument) as, e.g., can be inferred from the analysis 
of VIPERS passive galaxies shown by \citet{garilli2014}. As for the S/N, \ion{Ca}{2} 
$H$ and $K$ typically present an amplitude $F_{max}/F_{min}\sim3$, so in order to 
detect them a S/N$\sim$9 is required, which is also the typical S/N needed for a 
correct measurement of absorption features \citep[see][]{cappellari2009}.
This value represents a detecting threshold that allows the \ion{Ca}{2} $H$ and $K$ 
lines to be measurable; then the associated error (which, e.g., could be estimated 
with a MonteCarlo approach) will depend on the actual S/N of the spectrum. Once that 
it has been measured, then the contamination (and its corresponding uncertainty) 
of a sample can be obtained from Fig. \ref{fig:Hz}, and it could be propagated in 
the error analysis with Eq. \ref{eq:covy}.

The work-flow discussed above is meant to be applied on individual galaxies, and as 
a result the selection of galaxies without evidence of star formation (see e.g. 
criterion (2)) will depend on the properties and depth of the survey. However, the 
procedure can be further improved by considering stacked spectra \citep[as done 
e.g. in][]{moresco2012,moresco2016}, that will allow to significantly increase 
the S/N of the spectrum, and the detectability of spectral features. As an example, 
the stacking approach applied on BOSS data in \cite{moresco2016} allowed to increase 
the S/N of the spectrum by a factor between 10 and 60, demonstrating the robustness
of the selection by showing no evidence of emission lines even in the high-S/N 
stacked spectra. A more complete analysis of the detectability of features as a 
function of the quality of the spectrum will be addressed in a following work.

\vspace{5mm}

\section{Conclusions}
\label{sec:concl}
Motivated by the upcoming massive spectroscopic surveys that will 
significantly improve the statistics of massive and passive galaxies (especially 
at $z>1$), we have discussed how to select cosmic 
chronometers from such surveys. Because the spectra will 
be mostly in the optical rest-frame, we have focused on this region of the 
spectrum. 

The main systematics that could potentially affects the CC method are the 
dependence on stellar metallicity estimate, the reliance on SPS models, the 
progenitor bias, the presence of an underlying young component. In this analysis, 
we quantify the impact of a possible contamination due to a young component in two 
steps. Most importantly we provide the contribution to the covariance matrix due to 
this effect to be taken into account in the total error budget as a function of the 
contamination level. This is one of the novel contributions of this paper. Moreover, 
we provide new spectral indicators and illustrate a clear selection work-flow with 
which it is possible to minimize this contamination. This is also novel. While some 
of the individual steps were already discussed in the literature, their combination 
and integration was not, and is presented here for the first time. Our main results 
are summarized as follows:
\begin{itemize}
\item we identified the \ion{Ca}{2} $H/K$ indicator as a powerful diagnostic 
to identify the presence of a young underlying component.
This ratio, which in galaxies dominated by passive evolution is $>1$, shows an 
inversion already for a contribution of a young component (with ages $<200$ Myr) 
$\gtrsim$5\%.
Moreover, the analysis of previous data \citep{moresco2012,moresco2016} 
provided constraints of a contamination $\lesssim$1\%.
\item we analyzed the expected emission and absorption lines due to 
the presence of a young component, finding that it should present significant 
H$\alpha$ and [OII] emission at young ages ($\lesssim$10 Myr), or significant 
H$\delta$ absorption for a less recent star formation episode (hundreds of Myr up to 
$\sim$ 1 Gyr). The presence of a young component could also contribute to the rise of the 
UV flux, but in this case we estimated it to be significant (with fluxes comparable to the 
ones in the optical ranges) for ratios $\gtrsim$40\%; lower UV fluxes can be instead due 
either to a smaller young contribution, or to effects due to dust extinction. In this case, 
however, the NUVrJ diagram selection helps in disentangling possible age-extinction 
degeneracies.
\item we estimated the impact of a young component with different 
ratios $r$ with respect to the old one, on the $D_{n}4000$, finding a 
percentage variation of $\lesssim$5\% when $r<0.05$ (distinguishable from the 
$H/K$ inversion) up to $\gtrsim$40\% when $r\gtrsim0.4$.
\item finally, we propagated the previous effect, assessing its impact on
the measurement of $H(z$), providing a relation linking the systematic bias in 
$H(z)$ due to the presence of a young component to its percentage contribution 
$r$.
\end{itemize}
In particular, we demonstrated that the combination of these various 
indicators can be used to accurately discriminate the contribution of an 
underlying young population spanning a wide range of ages (0.01-1 Gyr). We 
also provide an explicit work-flow to reliably select the best candidates for 
CC. 
After this initial selection, excluding very young or localized star formation 
episodes, the resulting sample should undergo a further selection by mass and 
via a full spectral fitting to determine the (smooth) star formation history and 
the metallicity before proceeding with a $D_n4000$ analysis.

To assess the impact of the above effect on our previous analysis, we have re-analyzed 
the data presented in \citet{moresco2012,moresco2016} and show that 
indeed the young population contamination is minimal and consistent 
with zero given current uncertainties. We have calculated that at most it would 
bias the $H(z)$ determination by 0.4-1\% (at 1 $\sigma$, 0.8-2.3\% at 2$\sigma$), 
well below the current errors.
The fact that the diagnostic features are in the rest-frame optical 
spectrum, opens up the possibility of using future high-resolution spectroscopic 
surveys to yield an optimal ``golden" sample of CC.
Such sample can in principle provide constrains on $H(z)$ an order of magnitude 
better than current ones, provided that systematic errors can be kept under 
control. We have presented here a checklist and procedure to help in this 
direction for what a young stellar population contribution is concerned. 

A detailed study of the other sources of systematics, and how to 
minimize them, has been presented in \citet{moresco2012} and \citet{moresco2016}; 
let us just recall here that with massive spectroscopic surveys, the dominating 
effect is the one due to stellar metallicity estimate, and a future analysis on 
how to reduce it will follow.
This opens up the exciting possibility of constructing a sample of CCs 
where $H(z)$ can be measured at the \% level over the past 10 Gyr of cosmic 
evolution ($z\sim2$) thus testing the current $\Lambda CDM$ paradigm in a 
cosmology-model-independent way and searching for new physics.

\acknowledgments
We thank Adam Riess for useful comments and discussion. MM, ACim, LP and ACit acknowledge the grants ASI n.I/023/12/0 ``Attivit\`a relative alla fase B2/C per la missione Euclid'' and PRIN MIUR 2015 ``Cosmology and Fundamental Physics: illuminating the Dark Universe with Euclid''. We thank the anonymous referee for the comments, that helped to improve the presentation and clarify the discussion.
Funding for this work was partially provided by the Spanish MINECO under projects AYA2014-58747-P and MDM-2014-0369 of ICCUB (Unidad de Excelencia ‘Mar\'ia de Maeztu'). LV acknowledges support of European Union’s Horizon 2020 research and innovation programme ERC (BePreSySe, grant agreement 725327).

%

\vspace{5mm}


\software{BC16 \citep{bruzual2003}, M11 \citep{maraston2011}, Vazdekis et al. \citep{vazdekis2016}, CLOUDY \citep{ferland1998,ferland2013}, MOPED/VESPA \citep{heavens2004,tojeiro2007}}

\bibliographystyle{aasjournal} 
\bibliography{bib.bib}



\end{document}